\documentstyle[epsfig,endfloat]{ar2e}
\begin{document}
\baselineskip=15pt

\input epsf.tex

\input psfig.sty

\jname{Annu. Rev. Astron. Astrophys.}
\jyear{2004}
\jvol{42}
\ARinfo{1056-8700/04/0610-0000\$14.00}
%doi: 10.1146/annurev.astro.42.053102.134031

\setlength{\textwidth}{6in}
\setlength{\textheight}{8.5in}
\setlength{\oddsidemargin}{0.0in}
\setlength{\evensidemargin}{0.0in}
\def\rv{\revise}

\title{GRS 1915+105 and the Disc-Jet Coupling in
  Accreting Black Hole Systems}

\markboth{Fender \& Belloni}{GRS 1915+105 and Disc-Jet Coupling}

\author{Rob Fender
\affiliation{Astronomical Institute Anton Pannekoek, University of
  Amsterdam, Kruislaan 403, 1098 SJ Amsterdam, The Netherlands}
Tomaso Belloni
\affiliation{INAF---Osservatorio Astronomico di Brera, Via E. Bianchi 46, I-23807 Merate, Italy}}


\begin{abstract}
GRS 1915+105---the first stellar-scale, highly relativistic jet source
identified---is a key system for our understanding of the disc-jet
coupling in accreting black hole systems. Comprehending the coupling
between inflow and outflow in this source is important not only for
X-ray binary systems but has a broader relevance for studies of
active galactic nuclei and gamma-ray bursts. In this paper, we present
a detailed review of the observational properties of the system, as
established in the decade since its discovery. We attempt to place it
in context by a detailed comparison with other sources, and construct
a simple model for the disc-jet coupling, which may be more widely
applicable to accreting black hole systems.
\bigskip
\bigskip
\bigskip
\bigskip
\bigskip
\bigskip
\end{abstract}

\maketitle

\section{Introduction}
Jets---highly relativistic, collimated, powerful outflows---seem to be
an almost ubiquitous feature of accreting black holes, and yet remain
poorly understood. Widely studied in active galactic nuclei (AGN),
their association with the coupled process of accretion has been obstructed by the long
dynamical timescales associated with these objects. How
matter falling at high speed toward a black hole manages to
escape to the external universe remains a mystery {\footnote{To quote
Krolik (1999), ``In
principle one could imagine that accretion onto a black hole occurs
without any outflow at any point. Put another way, we do not know why
jets exist.''}} that can only be solved by
studying the coupled processes of inflow and outflow. The key to this
puzzle seems to lie with binary systems in our own galaxy, which
harbor stellar-mass (5--15 M$_{\odot}$)
black holes that---we are just beginning to
appreciate---seem to display the same kind of inflow:outflow coupling
as AGN but on timescales typically six to eight orders of
magnitude shorter in proportion to M. Thus, although AGN are the
powerhouses of the universe, it is their nearby
cousins,  which
vary rapidly enough that humans may follow the changes in the
disc-jet coupling, to which we turn for an understanding of
the relation between accretion and jets around black holes.

With the launch of the {\it Rossi X-Ray Timing} satellite ({\it RXTE}), an
unprecedented wealth of X-ray data from black hole candidates (BHC)
has become available to the astronomical community. In particular,
several bright transient systems have been followed with a large
number of densely packed observations (see McClintock \& Remillard
2004).  These observations have increased our knowledge of the
high-energy properties of these systems, although we still do not have
a complete picture of the physical mechanisms underlying
the X-ray emission.

GRS 1915+105, a binary system in our galaxy containing an accreting
black hole of probable mass $\sim 15$M$_{\odot}$, is a key object for
the study of the disc-jet connection. Its peculiar X-ray properties,
coupled with other characteristics in common with other systems, allow
us to study in detail the instabilities of the accreting gas. These
instabilities have been positively associated with the ejection of
relativistic jets and to related flaring episodes in the radio
band. This means that the interaction of emitted powerful
collimated jets can be studied on different timescales and associated
with (near-)simultaneous events in the accretion flow that provides
the gas for them. The uniqueness of GRS 1915+105 stimulated us to
center a full discussion on the disc-jet coupling on this system.

The review is structured as follows: In Section 2, the X-ray
and radio properties of GRS 1915+105 are described, followed by the
disc-jet coupling evidences in this system.  In Section 3, we describe
other BHCs in a similar fashion, whereas in Section 4,
we concentrate on discussing the comparison between all
systems. Our main conclusions are presented in Section 5.

\section{GRS 1915+105}

%  TMB section-----------------------------------------------------

\subsection{Discovery and Early Observations}

GRS 1915+105 was discovered on August 15, 2002, with the WATCH
instrument on board {\it GRANAT} as a 300 mCrab transient (Castro-Tirado et
al. 1992, 1994) and immediately detected at
higher energies by {\it CGRO}/BATSE (Harmon et al. 1992). A time-variable
radio counterpart (Mirabel et al. 1993a) and an IR counterpart
(Castro-Tirado et al. 1993, Mirabel et al. 1993b) were discovered in
the ROSAT HRI error box (Greiner 1993).  The X-ray spectrum appeared
to be variable and two-component (see, e.g.,  Alexandrovich et
al. 1994). {\it GRANAT}/WATCH and {\it CGRO}/BATSE monitoring of the source showed
periods of nondetection but never simultaneously in both
instruments, indicating that GRS 1915+105 was continuing the same
outburst started in 1992 (see Zhang et al. 1995, Sazonov \& Sunyaev
1995, Paciesas et al. 1996, Harmon et al. 1997). A BATSE hard X-ray
light curve extending up to the beginning of 1996 can be seen in
Figure 1. Paciesas et al. (1996) realized that the source
was showing an unusual level of variability, although the true
uniqueness of its nature was not clear from those
data.  The radio counterpart of GRS 1915+105 was observed to be
variable in flux and spectrum (see Rodr\'iguez \& Mirabel
1993), and the radio variability was correlated to the hard X-ray flux
(Mirabel et al. 1994).  The major breakthrough took place when Mirabel
\& Rodr\'iguez (1994) used the VLA radio to observe components moving
away from GRS 1915+105 with apparent superluminal velocity, at the
same time estimating a distance of 12.5 kpc (see Rodr\'iguez et
al. 1995), which excluded an extragalactic origin. GRS 1915+105 became
the first superluminal source in our Galaxy (see Section 2.3.1).

%COMP: Please insert Figure 1 here. Thank you

%\begin{figure}
%\centerline{\psfig{figure=harmon.ps,width=9cm}}
%\caption{BATSE 20-100 keV light curve of GRS 1915+105 in 5-bins. Flux
%conversion was done with a $\Gamma$=2.5 power-law model
%(from Harmon et al. 1997). Time ranges from mid-1991 to the launch of
%RXTE.}
%\label{batse}
%\end{figure}

\subsection{X-Ray Spectral Variability}

The launch of {\it RXTE} opened a new window to GRS 1915+105.  The large
area of the proportional counter array (PCA) (Jahoda et al. 1996), the
broad-band coverage provided by the High-Energy X-ray Timing Experiment (HEXTE) instrument (Rothschild et
al. 1998), the monitoring capabilities of the All-Sky Monitor (ASM)
(Levine et al. 1996), and the extreme flexibility of scheduling and
pointing were key factors that allowed a more complete view of
the unique X-ray properties of this system.  The full ASM light curve
from the start of the {\it RXTE} mission to October 12, 2003, averaged over
one-day bins, can be seen in Figure
2. These data start just after the time window of the BATSE data 
in Figure 1. From Figure 2, the peculiarity of the
variability of GRS 1915+105 on timescales longer than one day is evident.
Greiner et al. (1996) were the first to report an
unusual level of X-ray variability in GRS 1915+105 on timescales of
subseconds to days. The light curves presented in that work were complex
and structured in a way that was never observed before in any X-ray source.
The authors attributed this variability
to the effect of a major accretion disc instability and argued that major
events like those shown in Figure 3 could be related to
radio flares observed later.
A classification of spectral/timing patterns from early {\it RXTE}/PCA data was
presented by Chen et al. (1997), who identified two spectral states:
a hard one with a strong 0.5--6.0 Hz quasi-periodic oscillations (QPOs),
and a soft one with no QPOs (see also Paul et al. 1997).
In the same year, Belloni et al. (1997a) accumulated PCA energy spectra
from the high-flux and low-flux intervals seen in Figure 3
and fitted them
with the ``standard'' simplified model for BHCs, consisting of a
disc-blackbody component (see Mitsuda et al. 1984) plus a power-law
component to account for the high-energy tail. The free parameters of this
model are the flux $F_{pl}$ and photon index $\Gamma$ of the power-law and the
inner radius of the accretion disc $R_{in}$ and  temperature at the inner
radius $T_{in}$.
The strong and sudden hardening events
seen in the hardness curve in Figure 3 can then
be understood in terms of a hardening (from $\Gamma\sim$3.6 to
$\Gamma\sim$2.2) of the power law and a softening ($kT_{in}\sim$ 2.2
to $\sim$0.6 keV) of the disc component. At the same time, the inner-disc
radius $R_{in}$, which in many sources had been observed to remain
constant at a value of a few dozen kilometers (interpreted as the innermost
stable orbit, see Tanaka \& Lewin 1995; see below)
was seen to increase from $\sim$20 to $\sim$300 km. The interpretation of these
transitions, which happened on a timescale of a few seconds, was that
they were due to the onset of a thermal-viscous instability. In this framework,
the sudden increase in $R_{in}$ corresponded to the inner portion of the
accretion disc becoming unobservable in the X-ray band. The accreting matter
passing through $R_{in}$ would slowly ``refill'' the unobservable region
on a viscous timescale, increasing the surface density of the inner
portion of the disc until the reverse transition would take place and
go back to a fully observable disc. In this interpretation,
it is not necessary that the inner part of the disc is completely emptied
of gas, but simply that the gas makes a transition to a much cooler state,
which makes it effectively invisible in X-rays.

%COMP: Please insert Figures 2 and 3 here
%\begin{figure}
%\centerline{\psfig{figure=asm.eps,width=9cm}}
%\caption{RXTE/ASM light curve, in 1-day bins, from the start of the
%RXTE mission up to 2003 October 12.}
%\label{asm}
%\end{figure}

This picture was subsequently strengthened by the analysis of more observations
(Belloni et al. 1997b). Time-resolved spectroscopy allowed the evolution of the spectral parameters to be followed in detail (see Figure 4).
In particular, Belloni et al. (1997b) found that in the soft intervals,
$R_{in}$ was approximately constant around 20 km, independent of the length
of the interval, whereas in the hard intervals, the radius showed a time
evolution. The maximum radius reached in each hard interval was strongly
correlated with the length of the interval $t_{hard}$
(see Figure 5) and
anticorrelated with the minimum flux reached.
In the framework of the instability picture of Belloni et al. (1997a), this
has a direct interpretation:
The spectral evolution and the duration of the event are determined by
one parameter only, namely, the radius of the missing inner section of the
accretion disc, i.e., $R_{in}$. For a large radius, the drop in flux will be
large and the time needed for refill will be long.
Associating the refill time with the disc viscous timescale
$t_{visc}$ at $R_{in}$, can be expressed as

\begin{equation}
t_{visc} = 30 \alpha_{-2}^{-1} M_1^{-1/2} R_7^{7/2} \dot M_{18}^{-2}\, {\rm s},
\label{eq1}
\end{equation}

where $\alpha_{-2} = \alpha/0.01$, 
$R_7$ is the radius in units of 10$^7$cm,
$M_1$ is the black hole mass in solar masses, and $M_{18}$ is the accretion
rate in units of 10$^{18}$ g/s (see Belloni et al. 1997b). From this, under the reasonable assumption that the external $\dot M$ does
not vary within a one-hour observation, we expect to observe

\begin{equation}
t_{hard} \propto R_{in}^{7/2}.
\label{eq2}
\end{equation}

The line in Figure 5 shows a
fit of $t_{hard}$ versus $R_{in}$ with a $y=x^{3.5}$ relation: Only the longest interval deviates from the expected
relation. The two quantities $t_{hard}$ and $R_{in}$ are determined in
a completely different fashion, the first from timing analysis and the
second from a spectral decomposition, ensuring that the observed
correlation is not spurious.  The relation in
Figure 5 is a direct measurement of the radial dependence of a
fundamental quantity of an accretion disc, obtained through the
analysis of the variability from a source that is indeed
peculiar, but also shows many properties in common with other
systems. Owing to the approximations in
the disc blackbody model, the measurement of the inner-disc radius
cannot be considered precise in its absolute value (see Merloni et
al. 2000), although large variations, such as those observed in
GRS 1915+105, cannot be due to these approximations.  From
this observation, a one-to-one correlation between the length of an
instability interval and the following high-rate interval was found.
Yadav et al. (1999) found that indeed in many observations made with
the {\it IXAE} satellite, the length of the instability interval
correlated with the length of the following high-rate interval
and not with that of the preceding one. Therefore, the full observed high-low cycles are caused by the
onset of the instability, which is not directly influenced by the
duration of the ``stable'' interval.

%COMP: Please insert Figures 4 and 5 here

\subsubsection{Classification and Source Spectral States}

Markwardt et al. (1999), analyzing in detail a specific {\it RXTE}/PCA
observation from 1997, identified three separate states on the basis of 
timing and spectral characteristics. Figure 6 
shows the spectral characterization of these states in terms of
power-law photon index and inner-disc temperature.

%COMP: Please insert Figure 6 here

A complete classification of the light curves observed by {\it RXTE}/PCA from 
GRS 1915+105 during the first two years of the mission was presented
by Belloni et al. (2000). The observed light curves were divided into 
12 classes (shown in Figure 7) on the basis of their
timing and color properties. These 12 classes, later increased to 13
by Klein-Wolt et al. (2002) and possibly to 14 by Hannikainen et al. 
(2003), are not meant to be exhaustive but as a
starting point for subsequent analysis. From this classification and from a 
detailed analysis of X-ray color-color diagrams, Belloni et al. (2000)
concluded that, as suggested by Markwardt et al. (1999) for a
single observation, indeed the flux and spectral variability of GRS 1915+105 
can be interpreted as transitions between three separate states. They called
these states A, B, and C. Figure 8 shows a schematic view of the
three states in terms of their X-ray colors:

\begin{itemize}

\item State A: softer spectrum, dominated by a disc component
	with $kT_{in}>$ 1 keV. Mostly little time variability.
\item State B: softer spectrum, dominated by a disc component
	with a larger temperature than state A. Substantial red-noise
	variability on timescales $>$ 1 s.
\item State C: hard colors, with the spectrum dominated by a
	relatively flat power-law ($\Gamma$=1.8--2.5). White noise
	variability on timescales $>$1 s.
	
\end{itemize}

The transitions between these three states are so far observed only in 
the directions indicated by arrows in Figure 8, i.e., no direct
C$\Rightarrow$B are seen. These states connect to the instability picture
outlined above in the following way: state C corresponds to the onset
of the instability in the inner section of the disc, whereas states A and B
correspond to stable periods. This means that the transitions that involve
state A are not interpreted within that picture. 
The existence of definite events that cannot
be interpreted in a thermal-viscous instability framework was already 
pointed out by Taam et al. (1997), who analyzed a {\it RXTE}/PCA 
observation of class $\nu$. They showed that the difference between state
A and state B can be associated to a variation in the inner edge of the
accretion disc, although of smaller amplitude than that observed during
state C. This picture is confirmed by a more complete analysis by 
Migliari \& Belloni (2003).

%COMP: Please insert Figures 7 and 8 here

An analysis of time-resolved spectra from a large number of 
{\it RXTE}/PCA observations, together with timing analysis,  
was made by Muno et al. (1999), who presented a number of correlations
between spectral parameters and timing parameters. Vilhu \& Nevalainen (1998)
analyzed observations of class $\rho$ and
interpreted through spectral fitting that the regular ``ring''  in the color-color
diagram is due to out-of-phase oscillations
of different spectral parameters.

A full spectral analysis on a timescale of 16 s of the instability
oscillations in class-$\beta$ observations showed that indeed the inner-disc radius estimated from the disc blackbody component is larger
during state A than during state B (Migliari \& Belloni 2003). Moreover,
there is evidence for changes in the local accretion rate through the
disc during states C and B but not during state A, where inner-disc temperature
and radius change in a way that reflects a constant disc accretion rate (see
Figure 9).

%COMP: Please insert Figure 9 here

These results were followed by a series of theoretical papers aimed at a
more thorough exploration of the connection with accretion disc models
(Szuszkiewicz \& Miller 1998, 2001; Nayakshin et al. 2000;
Janiuk et al. 2000a,b; Nandi et al. 2001; Zampieri et al. 2001; Livio et al. 2003).

The complexity of the variability observed in GRS 1915+105 is unique and
difficult to model. Clearly, it will not be possible to interpret in detail
every aspect of the complex light curves shown in Figure
7, yet the structure of these light curves 
is not random and contains information about the accretion disc of
GRS 1915+105. As can be seen from different works (e.g., Belloni 2001),
there are complex light curves that repeat almost identically years apart.
Although at first sight the X-ray light curves observed from 
GRS 1915+105 seem wildly different, the number of repetition patterns 
observed by {\it RXTE} in the past nine years of observations is relatively small,
although clearly larger than 12. All the models listed in the previous
paragraph are aimed at explaining the single A--B--C event and its
appearing on different timescales, but the observed sequence of events
is complex and by no means random. This aspect has not been investigated
so far. For instance, Naik et al. (2002) found that certain variability
classes are observed preferably before and after long C-state 
intervals (plateaux, see below),
indicating a connection between the observed patterns and the long-term behavior, possibly linked to the mass accretion rate.

The issue of timescales deserves particular attention. As we have
shown above, the length of C-state intervals observed within one PCA
light curve can vary from a few seconds to half an hour. However,
there are observations when the source is observed only in state C
(defined as class $\chi$ in Belloni et al. 2000). Because each
observation lasts typically one hour, it is reasonable to hypothesize
that these are one-hour segments of longer state-C intervals. In some
cases, this can be measured directly. From Figure 2, one can
see intervals of relatively little variability in the daily ASM scans
(for instance, around days 200 and 650), corresponding to a higher
value of hardness ratio. These intervals have been called
plateaux. In addition, a longer, lower, extended hard state is
also visible in Figure 2, lasting more than a year and centered
around day 400. The color properties of these plateaux have been
analyzed by Belloni et al. (2000), who found that their properties are
consistent with being long state-C intervals. This is confirmed also
by timing analysis and by the radio properties (see below). A full
spectral analysis of the long plateau was presented by Trudolyubov et
al. (1999a), whereas Trudolyubov (2001) compared in detail the
properties of these three plateaux and found that both spectral and
timing characteristics differ between the short plateaux and the long
plateaux (see also Belloni et al. 2000). These differences are accompanied
by marked differences in the radio emission (see below).  Fast
variations, probably associated with B$\leftrightarrow$A state
transitions, have also been observed in class $\mu$
and $\rho$ (Taam et al. 1997, Belloni et al. 2000). They take place
on a timescale as low as 0.1 s.

Such a complex light curve as the one shown in Figure 2 has of
course been searched for patterns and periodicities. The problem is
complex owing to the fact that single ASM scans, lasting a few seconds, 
sample complex behaviors, such as those in Figure 7,
in a random way. A simple power density spectrum of this curve would not
yield believable results. Different approaches were followed.
Rau et al. (2003), on the basis of spectral analysis of
class-$\chi$ observations only, detected a 590-day-long term periodicity in
the power-law photon index,
which they also observe in BATSE and radio data, possibly related to 
disc precession. Greenhough et al. (2003), through differencing and
rescaling techniques that allowed them to separate different timescales, 
identify a 12--17 day timescale in the ASM light curve,
which might be related to the binary orbital period of 35 days.

An obvious question to be answered is why this complex behavior is so
unusual and has not been observed in any other sources. As shown below,
both (fast) timing and spectral properties of this system are not
at all peculiar. There is no definite answer as yet, although it is
widely assumed that the large value of mass accretion rate (which, depending
on the observation and on the spectral model, has been measured even to
exceed the Eddington luminosity for a 14$M_\odot$ object) plays a
fundamental role in setting up these instabilities.

\subsubsection{X-Ray Spectra}
In this section, we describe in more detail the X-ray energy spectra
that have been measured from GRS 1915+105 by different authors.

%COMP: paragraphs are 4th level heads

\paragraph{Broad-band spectra}

In the presence of strong variability on short timescales, it is not possible
to accumulate energy spectra with sufficient statistics to test models more
complicated than simple phenomenological approximations. Therefore, most
of the efforts have been directed to the fitting and interpretation of
data from observations with little variability.

Grove et al. (1998) reported {\it CGRO}/OSSE spectra (40 keV--10 MeV) of 
GRS 1915+105. In all three observations, a single power-law with no
observable high-energy cutoff up to 500  keV was observed, unlike other
BHCs in their low/hard state. OSSE observations span one
or two weeks: For the two observations of 1996 and 1997 during this period, the {\it RXTE}/PCA
database show observations of classes $\chi$ and $\mu$ in
1996 and of class $\alpha$ in 1997. This means that, although different states
were sampled by OSSE, GRS 1915+105 was probably in the C 
state for most of the time, when the power-law component is flatter. This was confirmed by
Zdziarski et al. (2001), who analyzed a total of nine OSSE observations.
The OSSE spectra, extending to 500 keV, do not show evidence of a 
high-energy cutoff. By checking the behavior of {\it RXTE}/PCA observations
taken during the OSSE periods, they see that at least one of these
observations consists probably of $\gamma$ behavior dominated by 
state B. The conclusion is that no high-energy cutoff is seen in either
state C or B (see Figure 10, color insert). The authors interpret this as evidence
for the presence of a hybrid population of thermal and nonthermal electrons
emitting through Comptonization and perform 
detailed joint {\it RXTE}/OSSE spectral
fits. Assuming isotropic emission, the resulting bolometric luminosity
is 6.5$\times 10^{38}$ and 1.7$\times 10^{39}$ erg/s for states C and B,
respectively. For a black hole mass of 14$M_\odot$, these values 
correspond to 0.35 and 0.93 of the Eddington luminosity.
Evidence for only a thermal Comptonization component was 
presented by Vilhu et al. (2001) on the basis of a sample of {\it RXTE} spectra.
These results are probably compatible, as the latter spectra did not extend
to 500 keV as the OSSE ones did.

%COMP: Figure 10 is color

The issue of the presence/absence of a high-energy cutoff is not easy
to solve.  BeppoSAX spectra corresponding to both B and C states were
analyzed by Feroci et al. (1999). The simultaneous 0.1--300 keV spectra
were fitted with a multicolor disc, a cutoff-power-law, and a Compton
reflection component.  The reflection was rather large, with a value
for $R=\Omega / 2\pi$ of $\sim$0.5 (where $\Omega$ is the solid angle
subtended by the reflecting matter to the illuminating source). A definite high-energy cutoff was found for all
observations, varying from 45 to 100 keV. Moreover, the cutoff energy
was positively correlated with the power-law photon index. These
results are clearly not compatible with the OSSE ones.  PCA and HEXTE
observations during the long plateau centered around day 400 in
Figure 2 led to the detection of a high-energy cutoff whose
value was rather constant around $\sim$50--100 keV (Trudolyubov et
al. 1999a). The model adopted here was again a simplified multicolor
disc plus cutoff-power-law. No correlation similar to the one reported
by Feroci et al. (1999) was seen.  As mentioned above, spectral
analysis of PCA observations during the long plateaux and the
shorter ones (days 200 and 650 in Figure 2), which are
associated to different radio properties (see below), showed that the
energy spectra are also different (Trudolyubov 2001). The shorter
plateaux have a much lower cutoff energy, as low as 12--20 keV. This is
in marked contrast with the OSSE results but is consistent with the
results of the color analysis of Belloni et al. (2000). A simple phenomenological model consisting of a multicolor disc
plus cutoff-power-law was used here again, to be compared with more complex
physical models used to fit the {\it RXTE}+OSSE spectra, which, however, were
not completely simultaneous given the length of the OSSE observations.
Fits to class-$\chi$ PCA+HEXTE observations over the first four years
of {\it RXTE} by Rau \& Greiner (2003) with a model consisting of a
reflected power-law (plus disc), featuring no high-energy cutoff,
yielded extremely high values for the reflection parameter $R$,
inconsistent with other measurements.

It is clear that the different spectral models used are mostly responsible for
the presence/absence of a high-energy cutoff. High signal-to-noise ratio
simultaneous data are necessary to resolve the issue. The shape of the high-energy component also affects the measurements
at lower energies, in particular, changing the normalization of the 
disc component, from which an estimate of the inner radius of the
optically thick disc component can be extracted.
Recent observations of GRS 1915+105 with INTEGRAL have been made and
preliminary results are now available. A simultaneous {\it RXTE} and INTEGRAL
observation during a plateau is reported by Fuchs et al. (2003b). Here
the energy spectrum was fitted to a simple power-law up to 400 keV, with 
no reflection component nor high-energy cutoff (see Figure 11, color insert).
Another INTEGRAL observation qualitatively confirms these results
(Hannikainen et al. 2003).

%COMP: Figure 11 is color

Overall, the broad-band 
energy spectrum of GRS 1915+105 can be described in terms
of three major components: a thermal component usually fitted with a
multicolour disc-blackbody, a hard component fitted with either a simple
power-law or a power-law with high-energy cutoff, or a more complex
Comptonization model (with or without Compton reflection) and a emission
line between 6 and 7 keV (see below). The relative contribution of these
components varies on different states. The spectra discussed
in this section are not different from those observed in other black hole
systems.

\paragraph{Spectral Lines}

A series of observations with ASCA between 1994 and 1999 produced the first
X-ray spectra with a resolution sufficiently high to explore narrow 
emission/absorption lines (Kotani et al. 2000). Absorption lines 
were detected and identified with resonant lines of calcium and iron, as 
well as blends of nickel and iron lines.
Similar lines were observed only in the other superluminal source
GRO J1655-40, and the authors speculated that they are associated to 
jet-production mechanisms.
A 30-ks observation with the Chandra/HETGS (Lee et al. 2002) confirmed the
presence of complex absorption structures, which were observed during a
probable C state.

Although at a lower spectral resolution, a BeppoSAX spectrum taken in 
1998 showed the clear presence of a broad emission line at 6.4 keV 
(Martocchia et al. 2002). The feature is broad and skewed, and fits
with a relativistic model indicated a nonzero spin for the black hole 
in the system (see Figure 12). 
These observations corresponded to an interval 
when the line was particularly intense out of a series of long BeppoSAX
exposures.

%COMP: Please insert Figure 12 here
%--------------------------------------------------------------------

\subsubsection{X-ray Timing}

As in the case of energy spectra, the fast timing features of
GRS 1915+105 are not as extraordinary as its variability on longer timescales. All observed properties in all three states appear rather
normal, as they have been seen in a number of other BHCs (see McClintock
\& Remillard 2004). For a definition and description of the different
components detected in the Fourier domain, see van der Klis (1995).

\paragraph{Noise and Low-Frequency Quasi-Periodic Oscillations}

Very complex power density spectra (PDS) have been presented by 
Morgan et al. (1997). Different continuum noise components and
QPOs are visible in the PDS, although the 
production of one PDS per observation, without separating different
intervals, makes it difficult to interpret the results. Three types
of QPOs were identified: high-frequency ($\sim$70 Hz) oscillations 
(described below); low-frequency (1-10 Hz) QPOs (see also Paul et al. 1997); 
and very-low-frequency QPOs, which can be identified with the state 
transitions described above.
A systematic difference between the PDS of soft and hard intervals was
identified by Chen et al. (1997). On their hard branch (corresponding
to state C, see above), the
PDS of GRS 1915+105 consists of a flat-top noise component plus a 1--10
Hz QPO. The centroid frequency of the oscillation is well correlated
with the count rate (see Figure 13) 
and shows higher harmonics that tend to disappear
at high count rates. In their soft branch (states A and B), the PDS is
steeper, with a broad peaked noise around 2 Hz. At high rates,
selecting state B intervals during oscillations, the peaked noise 
disappears and a broad 5-Hz QPO peak appears (Chen et al. 1997).
The three states identified by Markwardt et al. (1999) also have
distinct timing characteristics corresponding to those of Chen et al.
(1997).

%COMP: Please insert Figure 13 here

As in the case of spectral analysis, most of the effort has been 
dedicated to the C state, which is relatively easier to analyze because of the
long plateaux and the clear QPO at low frequencies (LFQPO).
Markwardt et al. (1999) found that the LFQPO, like similar oscillations in 
other systems, has an energy spectrum consistent with that of the hard
component, and that its centroid frequency correlates positively with 
the flux of the disc component (see also Muno et al. 1999). A correlation
between the minimum centroid frequency reached during a C interval and
the length of the interval (over a range of 5--500 s) was found by Trudolyubov
et al. (1999b), who showed that in about half of the cases this correlation 
is consistent with the relation between viscous timescale at a certain 
radius and Keplerian frequency at the same radius. Rodriguez et al. 
(2002a,b) presented a combined spectral/timing analysis of PCA data,
correlating the frequency of the QPO with the measured inner-disc radius,
whereas a correlation with the power-law photon index was investigated
by Vignarca et al. (2003).
Many groups concentrated their work on the PCA observations of class
$\chi$, when the source is always in state C. Many of these
observations are parts of longer plateaux (see above). Trudolyubov et
al. (1999a) analyzed PCA observations during the long radio-quiet
plateau of 1996--1997 (around day 400 in Figure 2) and
presented a number of correlations between timing and spectral
parameters. A thorough work on timing analysis of PCA observations
during both radio-loud and radio-quiet plateaux was done by Muno et
al. (2001), who considered the correlation between radio and timing
properties. Important differences in timing properties between
radio-quiet and radio-loud plateaux, associated to spectral
differences described above, was found by Trudolyubov (2001). During
radio-quiet plateaux, the total root mean square (rms) of the QPO is the same as during
radio-loud plateaux, but the properties of the noise component are
different. Its total fractional rms variability is
higher, and an additional broad component around 60--80 Hz is present
(see Figure 14, left panel).  Indeed, there seem to be two
separate ``flavors'' of C-state PDS: When the QPO frequency is at a
certain value, its width and integrated rms are always the same, but
the noise level can be significantly different (see also Figure
14, right panel). The association with radio flux differences
found by Trudolyubov (2001) is the first direct connection between
X-ray noise variability and radio properties.  The energy spectrum of
the oscillations is hard, with fractional rms increasing with energy,
but data from the {\it RXTE} HEXTE
showed a significant turnoff at high energies (30--40 keV; Tomsick \&
Kaaret 2001). Its first harmonic shows a decrease at 10 keV
(Rodriguez et al. 2002a).  The LFQPO shows significant and complex
phase lags. Reig et al. (2000) found that both QPO and continuum noise
show hard lags when the QPO frequency is below 2 Hz, evolving
smoothly to soft lags when the QPO frequency exceeds 2 Hz
(Figure 15; see also Lin et al. 2000). When the
fundamental frequency shows negative (soft) lags, the fist harmonic
shows positive (hard) lags (Tomsick \& Kaaret 2001).

%COMP: Please insert Figures 14 and 15 here

Little work has been done on the timing properties of states A and B.
Power spectra were shown by Chen et al. (1997), Markwardt et
al. (1999), and Belloni (1999a). As mentioned above, the PDS of the
soft states are complex and lack clear narrow QPO peaks like those
from state C. A set of PDS from variability classes $\phi$, $\nu$, and
$\delta$ (see Figure 7) have been presented by Ji et
al. (2003). These authors also performed a coherence analysis and found
that the coherence deviates from unity above a characteristic
frequency. A few PDS are also presented by Reig et al. (2003),
selected to separate state A and state B.

\paragraph{High-Frequency Quasi-Periodic Oscillations}

>From a set of 31 early {\it RXTE}/PCA observations of GRS 1915+105, Morgan et al.
(1997) discovered the first QPO at frequencies $>$ 50 Hz in a BHC.
In only two observations of those analyzed, a significant peak at a frequency
of 67.7 and 65.5 Hz was detected, with a width of $\sim$4 Hz and
a fractional rms variability of 1.1 and 1.6\%, respectively. Its
rms increased with energy up to 6\% above 12 keV. Hints of the presence
of the same oscillation were present in four more observations.
Belloni et al. (2001) analyzed the observation when the QPO was strongest
(May 5, 1996) and found strong variations in its energy dependence as
a function of the much slower B$\leftrightarrow$A oscillation. In a second
observation, they found a broad 27 Hz QPO only during the A intervals.
The phase lags of the 65 Hz QPO were studied by Cui (1999), who found
hard lags increasing with energy up to 2.3 radians between the 5.2--7.0 keV
and the $>$13 keV bands.

Strohmayer (2001) analyzed 97 PCA observations from 1997 and found in 5
of them a QPO at a frequency ($\sim$69 Hz) higher than that found by
Morgan et al. (1997). In addition, a second QPO peak at $\sim$40 Hz was
found in the high-energy data, simultaneous with the 69 Hz peak
(see Figure 16).

%COMP: Please insert Figure 16 here

These oscillations were found in observations of classes $\gamma$ and $\phi$
and are therefore associated with states B and A, but not with state C, when
the low-frequency QPOs are observed. It is interesting that the three 
frequencies presented above, 27, 40, and 69 Hz, are roughly in 2:3:5 ratio
(see Abramowicz et al. 2003).

A representative set of PDS from the three different states 
is shown in Figure 17 (from Reig et al. 2003).

%COMP: Please insert Figure 17 here

%  RPF section-----------------------------------------------------

\subsection{Radio and Infrared Emission}

\subsubsection{Radio}

GRS 1915+105 has turned out to be one of the most spectacular radio
sources within our galaxy, displaying not only---like in X-rays---highly variable and structured light curves but also spatially
resolvable steady and transient jets at different times.

\paragraph{Radio variability and relativistic jets}
Mirabel et al. (1993a) reported the first detection of the radio
counterpart of GRS 1915+105 as a relatively weak (2.5--5 mJy) source
at 20 cm with the Very Large Array (VLA).  Although interesting, there
was little hint in this first IAU Circular of the excitement to
follow. However, by the end of 1993, Mirabel et al. (1993b) and
Rodr\'iguez \& Mirabel (1993) reported further VLA observations,
indicating a highly variable, sometimes very luminous, and possibly
even moving radio source.

Early in 1994, Mirabel et al. (1994) published details of the radio and
IR counterpart. They demonstrated that the source probably lies
behind more than 30 magnitudes of absorption in the optical band---which has to date precluded any identification of an optical
counterpart (but see Bo\"er, Greiner \& Motch 1996 for a possible
I-band counterpart)---and that the radio counterpart showed
variability that was in some way correlated with the X-ray
emission. In March 1994, Gerard, Rodr\'iguez \& Mirabel (1994)
announced a major radio outburst (flux densities $>$ 1 Jy at 20cm) from the
source. It was this outburst that was to change the way we look at
jets from X-ray binaries: In a ground-breaking paper, Mirabel \&
Rodr\'iguez (1994) reported apparent superluminal motions in twin-sided
radio knots moving away from the core of GRS 1915+105 (Figure 18, left
panel). For the first time, superluminal motions---previously a
characteristic unique to jets from supermassive black holes in active
galactic nuclei---had been observed in our galaxy. The term
microquasar, first applied to the galactic X-ray source 1E
1740.7-2942 owing to its large-scale radio lobes (Mirabel et al. 1992),
seemed entirely appropriate for GRS 1915+105. The fact that powerful,
significantly relativistic ejections could be observed from a black
hole transient that, although different in some ways from other
transients was not that different, ushered in a new era in the
study of jets from stellar mass compact objects. The
nonuniqueness of such highly relativistic jets was dramatically
highlighted by the discovery of a superluminal jets from another X-ray
transient, GRO J1655-40, within one year (Hjellming \& Rupen 1995, Tingay et al. 1995). 

%COMP: Please insert Figures 18 and 19 here

Rodr\'iguez et al. (1995) reported five months' radio monitoring of
GRS 1915+105, which demonstrated that the source was undergoing
repeated major radio flares.  Radio monitoring had also begun at the
Green Bank Interferometer (GBI) (Foster et al. 1996) and the Ryle
Telescope (RT) (Pooley \& Fender 1997). Foster et al. (1996) reported
detailed GBI monitoring at 2.3 and 8.3 GHz, which identified two
states of bright radio emission: flaring (optically
thin spectra, rapid decays) and plateaux
(flat-topped periods in the light curve with optically thick
radio spectra). We consider the
plateaux to be a bright subset of the $\chi$ class/C state of GRS
1915+105 (also refered to as $\chi_{\rm RL}$ by Naik \& Rao
2000 and type II hard
states by Trudolyubov 2001). A clear example
of a plateau state is presented in Figure 19.

That the flaring periods corresponded to further relativistic
ejections was assumed and confirmed by Rodr\'iguez \& Mirabel (1999),
who also noted small apparent changes in the projection of the jets on
the sky.  Later the same year Fender et al. (1999) reported
observations of a major outburst with the UK Multi-Element Radio
Linked Interferometer Network (MERLIN), with five times the angular
resolution of the VLA at 5 GHz (Figure 18, right panel). These
observations revealed again relativistic ejections, but with
considerably higher (by approximately 20\%) proper motions than those measured
with the VLA. Comparison with GBI and RT radio monitoring, and X-ray
monitoring with the {\it RXTE} ASM, indicated that the multiple relativistic
ejections observed by Fender et al. (1999) orginated in major flares
immediately following a protracted plateau phase. The MERLIN
observations further revealed significant and variable (possibly
rotating) linear polarization in the ejecta, and hints of curvature in
the jet axis, possibly owing to precession.

The beauty of the observations of two-sided proper motions by Mirabel
\& Rodr\'iguez (1994) and Fender et al. (1999) is that they allow
several clear results to be drawn. A maximum distance, under the
assumption of an intrinsically symmetric ejection, can be derived as

\[
d_{\rm max} = \frac{c}{\sqrt(\mu_{\rm app}\mu_{\rm rec})},
\]

where $\mu_{\rm app}$ and $\mu_{\rm rec}$ are the approaching
and receding proper motions,respectively, and $c$ is the speed
of light. The proper motions reported by Mirabel \& Rodr\'iguez
(1994) indicated that GRS 1915+105 lay within our galaxy and
was therefore an X-ray binary source. The most recent two-sided proper
motions, reported in Fender et al. (1999), constrain $d_{\rm max} \leq
13.6$ kpc (3$\sigma$). The observed proper motions can also be used to
place a lower limit on the velocity of the radio-emitting knots,
independent of distance:

\[
v_{\rm min} = \frac{\mu_{\rm app}-\mu_{\rm rec}}{\mu_{\rm
app}+\mu_{\rm rec}} c,
\]

which, when applied to GRS 1915+105, indicates a minimum velocity of
$\sim 0.3c$ on VLA scales and $\sim 0.4c$ on MERLIN/Very
Long Baseline Array (VLBA) scales (see below). The discussion of what
can be derived from the proper motion measurements is expanded upon in
Mirabel \& Rodr\'iguez (1999) and Fender (2003).

Mirabel \& Rodr\'iguez (1994)
also noted that the flux ratio between the
approaching and receding components was very close to that expected in
the case of true bulk motion:

\[
\frac{S_{\rm app}}{S_{\rm rec}} = \left(\frac{1+\beta \cos
\theta}{1-\beta \cos \theta}\right)^{k-\alpha},
\]

where $\beta = v/c$ and $\theta$ are the intrinsic velocity and angle
to the line of sight of the jets, $\alpha$ is the spectral index of
their radio emission, and $k$ ranges between 2 (for a steady flow) and
3 (for discrete ejections). The observed flux ratios of $8 \pm 1$
found by Mirabel \& Rodr\'iguez (1994) and 6--10 by Fender et al. (1999)
correspond to a value of $k$ in the range 1.3--2.3 and indicate true bulk
relativistic motions at some level. This does not rule out an origin
for the observed knots in moving shocks but does require that the
underlying plasma flow has bulk relativistic motion. In more detailed calculations, Bodo \&
Ghisellini (1995) concluded that the
observed patterns could be shocks but still required a bulk motion of
$\geq 0.7c$ (see also Atoyan \& Aharonian 1999).

The power in these ejections was also found to be very
large---applying the minimum energy conditions of Burbidge (1959),
Mirabel \& Rodr\'iguez (1994) estimated an energy associated with
the March 1994 event of $3 \times 10^{46}$ erg.  Conservatively estimating an injection/acceleration
timescale of $\leq 3$ days, this corresponds to a jet power during
this period of $\geq 10^{41}$ erg s$^{-1}$. Atoyan \& Aharonian (1999)
and Fender et al. (1999) estimated lower, but still considerable, jet
powers of $\geq 10^{38}$ erg s$^{-1}$.  Furthermore, as demonstrated
in Fender (2003), the observed proper motions only allow us to place a
lower limit on the bulk Lorentz factor $\Gamma$ of the flow, and
therefore on total energy associated with the ejections (the total
energy $E_{\rm tot} \sim (\Gamma-1)E_{\rm int}$, where $E_{\rm int}$
corresponds to the above minimum energy values). The true value of the
initial bulk Lorentz factor for the relativistic ejections, such as
those observed by MERLIN, is likely to lie in the range

\[
2 \leq \Gamma_{\rm bulk} \leq 30,
\]

where the lower limit can be derived from observations and the upper
limit corresponds to a mean jet power of $\sim 10^{41}$ erg s$^{-1}$
over a 12-h jet formation phase (at $\sim 45$ times the Eddington
limit for a $\sim 15$ M$_{\odot}$ black hole, this should be
reasonable!). Clearly, the power expended on particle acceleration and
bulk motion during the jet formation episodes could represent a
significant, even dominant, fraction of the total accretion power
available.

The nature of the compact self-absorbed emission during plateaux was
revealed by Dhawan, Mirabel \& Rodr\'iguez (2000) who found in
observations with the VLBA that it
correponded to a compact, relatively steady jet.  They further reported
proper motions on small scales that were consistent with those
reported by Fender et al. (1999). The lower values for the proper
motions reported in VLA observations (Mirabel \& Rodr\'iguez 1994, 1999) may be due to resolution effects or genuine
deceleration of the ejecta as they propagate further from the launch
point. Further very long baseline interferometry (VLBI) observations with the European VLBI Network (EVN)
have confirmed the resolved core during plateux states (Giovannini et
al. 2001) and suggested that the plateaux jet is of lower velocity
than the transient ejections, although the reported details were
sparse. The most recent dual-frequency VLBA observations (Fuchs et
al. 2003b) reveal a compact jet from the source whose size varies as a
function of frequency in a way that appears to be consistent with
expectations for a self-absorbed outflow (Figure 20).

Shortly after the major relativistic ejections were reported, several
groups realized that these jets might leave their imprint on
the ISM, just as the jets of AGN produce bright radio lobes. Levinson
\& Blandford (1996), Kaiser et al. (2000, 2004), Heinz (2002), and others have discussed the formation of shocks in the ISM by
these jets. However, although large-scale jet-powered
nebulae are known around some X-ray binaries (see, e.g., Fender 2004
and references therein), searches for large-scale radio structures
around GRS 1915+105 have been generally unsucessful (Chaty et
al. 2001, Ostrowski \& F\"urst 2001). Nevertheless, there are hints of
associations with nearby IRAS sources that deserve further
investigation (Chaty et al. 2001; Kaiser et al. 2004). Note though
that these sources are so far distant that if associated it must have
been with outbursts much earlier than the one that has
been ongoing for the past decade (Chaty et al. 2001 and discussion
therein).

\paragraph{Radio ``oscillations''}
Another major discovery was made with the RT: Pooley (1995)
  reported ``oscillations'' at 15 GHz with an apparent period close to
  38 min.  Monitoring with both instruments continued apace, and in
  1997, Pooley \& Fender (1997) reported comprehensive long-term radio
  monitoring at 15 GHz. The presence of quasi-sinusoidal oscillations
  (see also Rodr\'iguez \& Mirabel 1997), with amplitudes up to 50
  mJy and periods in the ranges tens of minutes to hours, was
  unambiguous. Furthermore, Pooley \& Fender (1997) demonstrated that
  oscillations at 15 GHz led those at 8.3 GHz by
 approximately 4--5 min. Finally, they showed
  that there appeared to be some connection between these radio
  oscillations and X-ray ``dips'' observed at the same time, with a
  similar recurrence period, by {\it RXTE} (Figure 21). This was the first example of the great
  successes that were to come by combining {\it RXTE} with
  simultaneous radio and/or IR observations.  Fender et
  al. (1999) noted that approximately four days of radio
  oscillations separated two major ejection events resolved with
  MERLIN. Naik et al. (2001) have
  also discussed the association of dips in the X-ray light curve with
  radio flaring.

More recently, Klein-Wolt et al. (2002) have reported a comprehensive study
of the relation between radio and X-ray emission in GRS 1915+105 (see
below for more details). Fender et al. (2002a) reported a detailed
study of the radio oscillations from GRS 1915+105, revealing that the
delays between two frequencies are not constant from epoch to epoch.
Finally, Fender et al. (2002b) reported the detection of circular
polarization associated with relativistic ejections from GRS 1915+105
utilizing simultaneous ATCA and MERLIN observations.

\subsubsection{Infrared and Millimeter}

\paragraph{Infrared Variability}
Mirabel et al. (1993b, 1994) first identified the
near-IR counterpart of GRS 1915+105 as a variable
source in the J (1.2 $\mu$m), H (1.6 $\mu$m), and K (2.2 $\mu$m)
bands. Whereas spectroscopy and high-resolution imaging
(see below) produced some surprises, a key piece in the disc-jet
connection came from high time-resolution IR photometry
of the source. Mirabel et al. (1996) reported a brightening and
reddening of the source in the near-IR following a
major ejection event, which they interpreted as reverberation of the
event in circum-binary dust. Chaty et al. (1996) reported more rapid
near-IR variability. Motivated by this and the
discovery of radio oscillations (see above)
(Pooley 1995, Pooley \& Fender 1997), Fender et
al. (1997) performed relatively high time-resolution K-band photometry
of GRS 1915+105 and found the source to be oscillating with a
comparable amplitude (when dereddened), period, and shape as radio
oscillations observed earlier the same day. They argued that the
IR events, like the radio events, were likely to be
synchrotron in origin, thereby increasing at a stroke the minimum
power associated with these oscillation events by four to five orders
of magnitude.  Fender et al. (1997) further
suggested that IR synchrotron emission could explain
the result of Mirabel et al. (1996) (see above) without requiring dust
reverberation.  Like the major ejections (see above), these smaller,
repeated ejections were clearly a major power output channel.

Follow-up to these observations was intense and rapid. Eikenberry et
al. (1998a) reported high time-resolution IR photometry
simultaneous with {\it RXTE} observations, confirming the suspected
link between events (Figure 22).  Mirabel et
al. (1998) observed simultaneously in the radio, IR,
and X-ray bands (Figure {23}). These observations appeared to reveal a
hard X-ray dip followed by an IR flare, followed by a radio flare, thereby revealing a progression
of the event to lower frequencies, as expected for models with
decreasing optical depth in ejecta. Eikenberry et al. (1998b) reported
IR spectroscopy during flare/oscillation events, which
revealed that the IR emission lines varied in strength
in an approximately linear way with the continuum. The mechanism
behind this is not immediately obvious because it was
already concluded that the continuum was (mostly) synchrotron emission
in origin. Eikenberry et al. (1998b) suggested that the lines were
being radiatively pumped by the synchrotron flares. Fender \& Pooley
(1998) presented further simultaneous radio and IR
observations, revealing a comparable IR-radio delay to
that reported by Mirabel et al. (1998). Ogley et al. (2000) reported
``excess'' submillimeter emission from GRS 1915+105, and
Fender \& Pooley (2000) reported simultaneous IR and
millimeter (1.3 mm) observations of GRS 1915+105, revealing
large oscillation events taking place at both frequencies near
the end of an apparently uninterruped sequence of over 700
events. Eikenberry et al. (2000) presented evidence for weaker
IR flaring from GRS 1915+105 that was
associated with soft dip/flaring cycles.

%COMP: Please insert Figures 22 and 23 here

At longer wavelengths, Winkler \& Trams (1998) and Fuchs et al.
(2003a) report ISO observations of GRS 1915+105. One of the
observations of Fuchs et al. (2003a), between 4--18 $\mu$m, was during
a plateau state and is consistent with a flat-spectrum component,
probably synchrotron or free-free emission (see below).

\paragraph{Spectroscopy} 
Castro-Tirado, Geballe \& Lund (1996) 
reported K-band spectroscopy of
the source, revealing emission lines of HI, HeI, and HeII. Whereas
Castro-Tirado et al. (1996) and Eikenberry et al. (1998b) argued that
the emission lines probably arose in an accretion disc in a low-mass
X-ray binary, Mirabel et al. (1997) argued that the lines were likely
to originate in an Oe- or Be-type massive star and that GRS 1915+105
was therefore a high-mass X-ray binary.  The origin of these lines
remained controversial until the identification of the binary
companion by Greiner et al. (2001a,b) (see below). 

\paragraph{Resolved Infrared Jets} 
Sams, Eckart \& Sunyaev (1996a) reported the discovery of a
near-IR jet in K-band speckle imaging of GRS 1915+105 with the
ESO New Technology Telescope (NTT). Repeat observations less than one
year later revealed that the jet disappeared (Sams, Eckart \&
Sunyaev 1996b). Eikenberry \& Fazio (1997) also found no evidence for
extended IR emission from GRS 1915+105 and, although the results of
Sams, Eckart \& Sunyaev (1996a)
appear convincing, it is odd that the IR
extension has never been independently verified.

\paragraph{Identification of the Orbit, Companion, and Mass Function}
Greiner et al. (2001a) finally managed to identify the binary companion to GRS
1915+105 by observing $^{12}$CO and $^{13}$CO bandheads in H- and
K-band spectroscopy using ESO's Very Large Telescope (VLT). These
features clearly identified the binary companion as a K- or M-type
giant, confirming that GRS 1915+105 was, in fact, a low-mass X-ray
binary. More excitingly, photospheric absorption features such as
these opened the possibility to perform radial velocity studies of the
system to see if, as expected, it contained a black hole. This is
precisely what Greiner, Cuby \& McCaughrean (2001b) did: In a
comprehensive IR spectroscopic study, they obtained an
orbital period of $33.5$ days with an associated radial velocity
semi-amplitude of $140 \pm 15$ km s$^{-1}$ (Figure {24}). The
resulting mass function (minimum mass for the accreting object) is
$9.5 \pm 3.0$ M$_{\odot}$, indicating a black hole at the $\sim 2.5
\sigma$ level. Assuming a lower-limit on the mass of the binary
companion of 1.2 M$_{\odot}$ and an orbital inclination corresponding
to that derived for the relativistic jets by Mirabel \& Rodr\'iguez
(1994) of $70 \pm 2$ degrees, they arrived at a most likely mass for
the compact object in GRS 1915+105 of $14 \pm 4$ M$_{\odot}$, making
it a strong candidate for the most massive stellar
mass black hole known. Several authors have considered the
evolution of the GRS 1915+105 binary system in  light of
this result (e.g., Belczynski \& Bulik 2002; Podsiadlowski,
Rappaport \& Han 2003 ; see also, e.g., King 2004 for a comparison with
ultra-luminous X-ray sources in external galaxies).  It is
important to realize that the binary plane may not be
perpendicular to the jet axis (Maccarone 2002), in which case the mass
of the black hole in GRS 1915+105 could be significantly greater
(although not much smaller) than the estimate of Greiner, Cuby \&
McCaughrean (2001b).

%COMP: Please insert Figure 24 here

\subsection{Radio and Infrared Evidence for Disc-Jet Coupling}

\subsubsection{First Connections from Daily Monitoring}

The first obvious connections between the disc (X-rays) and jet (radio
initially; later the jet was also appreciated to emit in the
near-IR), were identified via daily monitoring of the source in
multiple bands.

Foster et al. (1996) and Harmon et al. (1997) first demonstrated the
clear association between periods of hard X-ray emission and major
radio outbursts in GRS 1915+105. Based the imaging of superluminal
ejections associated with just such an outburst by Mirabel \&
Rodr\'iguez (1994), it was reasonable to make a connection between these
long, hard X-ray phases and major ejections. Foster et al. (1996) noted
that the radio emission during the hard X-ray phase was actually
rather optically thick and coined the phrase plateau states to
describe this combination of hard X-ray emission and strong
flat-spectrum radio emission. The association with radio flares and
ejections was---as discussed below---probably due to the fact
that such plateaux generally reverted to softer X-ray states within
10--30 days, and that this transition was accompanied by a major
ejection event (Fender et al. 1999, Klein-Wolt et al. 2002).  Figure
19 presents multiwavelength radio and X-ray monitoring of
GRS 1915+105 through such a plateaux phase.

\subsubsection{Coupling between X-ray ``Dips'' and Synchrotron Oscillations}

%Mirabel et al. (1998) suggested that a `spike' in the X-ray light
%curve, at which point the X-ray spectrum softened considerably, was
%the point at which the infrared began to rise and was therefore an
%indicator of the `launch' of the ejection (see Fig {\ref{m98}}).

By 1998 it was already established that periods of ``hard dips'' in the
X-ray light curve of GRS 1915+105 on timescales of minutes to hours
were associated with events in the IR and radio bands. Following
the initial hints in Pooley \& Fender (1997), key observations were
reported in Eikenberry et al. (1998a,b) and Mirabel et al. (1998).
More recently, Klein-Wolt et al. (2002) have undertaken the most
thorough study of the connection between the X-ray variability and
radio emission, and they highlight a strong link between hard X-ray states
and radio emission, and specifically a one-to-one correspondence
between hard X-ray (state C) ``dips'' and radio oscillation events.

It seems clear that, although the details of the physics may not be
well understood, each hard dip in a sequence is associated with a
radio---millimeter---IR event, which itself is associated with a
powerful outflow. Note that Feroci et al. (1999) have reported the
association of a radio ``flare'' from GRS 1915+105, associated with a
dip in the X-ray light curve; with hindsight this event seems more
like an oscillation event associated with an isolated state C. The
key observational characteristics of these oscillation events are as
follows:

\begin{itemize}
\item{Each event is associated with a state C dip followed by a state A/B
phase.}
\item{The rise and decay profiles are very similar over a broad range in frequencies (from the radio through the near-IR). This suggests that the optical depth effects do not strongly influence the profile of the event, and that adiabatic expansion losses (which are frequency-independent) dominate the decay phase.}
\item{The events occur with a delay between different frequencies in the sense that higher frequency events occur earlier; the delays are not constant to better than a factor of two. These delays suggest optical depth effects.}
\item{The energy associated with the events is large, and it is a significant fraction of the contemporaneous X-ray luminosity.}
\end{itemize}

Whether these synchrotron events correspond to the ejection of
discrete blobs (e.g., Mirabel et al. 1998), internal shocks in a more
continuous flow (Kaiser, Spruit \& Sunyaev 2000) or variations in the
jet power in a self-absorbed outflow (Collins, Kaiser \& Cox 2003) is
currently not clear (but see below).  That are powerful events --
extracting a sizeable fraction of the available accretion energy
(assuming it is Eddington limited) seems indisputable (see, e.g., power
estimates in Fender \& Pooley 2000).

\subsubsection{``Steady'' Jets in Prolonged Hard X-ray Plateaux}

Foster et al. (1996) discovered that prolonged
phases of hard X-ray emission---plateau states---were associated
with steady, relatively bright, flat-spectrum radio emission,
indicating some (self-)absorption of the synchrotron emission.  Dhawan
et al. (2000) and Fuchs et al. (2003b)
have reported direct imaging of
an apparently steady radio jet during this phase (Figure 20).
The nature of this steady radio emission has been discussed at length
in, e.g., Falcke \& Biermann (1996), Muno et al. (2001), Klein-Wolt et
al. (2002), and Vadawale et al. (2003).  We can now be confident that
each plateau episode is associated with such emission, which can
naturally be explained in terms of a conical self-absorbed jet
(e.g., Blandford \& Konigl 1979, Hjellming \& Johnston 1988, Falcke \&
Biermann 1996).

\subsubsection{Major Ejections After (and Before?) Plateaux}

Although empirically well established, the physical connection between
X-ray emission and the most spectacular of the jet-related phenomena---the superluminal jets---has been the hardest to understand. This
is perhaps not surprising given that they do not repeat
quasi-periodically many times in a row (like the oscillation events),
nor are they quasi-steady for many days (like the plateaux).

However, a general characteristic did emerge early on, namely that
plateau states are generally followed (and perhaps preceded) by major
ejection events (e.g., Foster et al. 1996, Fender et al. 1999, Dhawan
et al. 2000, Klein-Wolt et al. 2002) (see Figure
19). Furthermore, it seems rather common that the ``post
plateau flare'' (Klein-Wolt et al. 2002) is followed by several days of
X-ray/radio oscillations. Klein-Wolt et al. (2002) also noted that
the pre-plateau flares are generally stronger than the post-plateau
flares.  In the study of Fender et al. (1999), these oscillations
appeared for four days between major relativistic ejection
events.

\subsubsection{The Overall Picture---Two or Three Types of Jets?}

Above, we have outlined the observational characteristics of
three types of radio emission associated with jets from GRS 1915+105:
oscillations, plateaux, and major flares. That the source can readily
switch between these modes is evident from the data presented in
Fender et al. (1999) and Klein-Wolt et al. (2002) wherein the source

\begin{enumerate}
\item{Produces a medium-strength optically thin flare (probable ejection event)}
\item{Enters a plateau state (steady jet)}
\item{Produces a major flare (discrete relativistic ejection)}
\item{Enters several days of oscillations (repeated small ejections)}
\item{Reflares several times (discrete relativistic ejections)}
\end{enumerate}

We have no reason to think that this behavior was particularly
   unique (inspection of the radio light curves in, e.g., Pooley \&
   Fender 1997 suggest that it definitely is not).

How many types of jet production are required to explain this
phenomenology? The answer would seem to be two or three.

The steady plateau jets and the major ejections are clearly different,
as outlined in Table {\ref{jetcomp}}.

%COMP: PLEASE INSERT TABLE 1 HERE
\begin{table}
\caption{Comparison between the properties of the two apparently
distinct types of jets produced by GRS 1915+105}
\begin{tabular}{lr}
\toprule
Plateaux jets & Relativistic ejections \\
\colrule
Can be steadily produced for & Produced in $<24$ h \\
up to $\sim 20$ days         & \\
\colrule
Optically thick spectrum & Rapidly evolve to an \\
                         & optically thin spectrum \\
\colrule
Low polarization ($<5$\%)& Up to 20\% polarization \\
\colrule
Some evidence for low velocity & Highly relativistic ($\Gamma >2$) \\
\colrule
Associated with steady hard X-ray & Associated with
hard$\rightarrow$soft \\
(state C) phases                  & X-ray state changes \\
\botrule
\label{jetcomp}
\end{tabular}
\end{table}

It is possible that the oscillation events correspond to a third
class of jet formation, but it is likely that they do not.
Based on the empirical connections established in Klein-Wolt et
al. (2002), we can see that the oscillations are in fact associated
with a combination of steady state C phases and with transitions
from this state C to a softer state.  Note that Naik \& Rao (2000) and
Naik et al. (2001) interpret sequences of dips/oscillations as the cause of the major flare events, although this seems at
odds with observations where many days of oscillations have not been
associated with major flaring (e.g., Pooley \& Fender 1997, Fender \&
Pooley 2000).

We therefore conclude---being consistent with Occam's
razor---that the
oscillation events are the equivalent of the optically thin events
that follow plateaux. We shall see in the next section
that this interpretation is supported by drawing analogies with other
objects. One obvious consequence of this interpretation is that,
regardless of the velocity of the steady plateaux jets, the
oscillation events should also be associated with relativistic bulk
velocities.

\section{Other Sources: GRS 1915+105 in Context}

\subsection{Black Hole X-ray Binary States}

Some of the observed X-ray and IR/radio properties of GRS
1915+105 are clearly unique to this system. On the other hand, many other properties are similar to those observed
from other black-hole binaries. In this Section, we briefly
summarize the general common properties of other systems,
leaving out other peculiarities, to compare them with those
of GRS 1915+105. We do not discuss the low-luminosity quiescent
state because GRS 1915+105 has been very bright since its discovery,
not allowing a comparison with quiescent systems.

Even after years of {\it RXTE}  data from both the (few) persistent systems and
the (many) transient systems, our general picture of the X-ray properties
of BHCs is not much clearer in the sense that there are a number of
significant exceptions to the states-scheme discussed below. Also on the
theoretical side, there is not much consensus as to the physical origin 
of some of the observed spectral components, the timing feature, nor
 the physical parameters driving state transitions. Yet, a classification 
in terms of source states is very useful and our only current way to 
provide an observational view for the development of theoretical models.

Currently, we identify three main ``canonical'' states in BHCs,
identified by their timing and spectral properties. We
summarize here their basic properties; for a complete description see,
e.g., Tanaka \& Lewin (1995), McClintock \& Remillard (2004), J. Homan \&
T. Belloni (manuscript in preparation).

\subsubsection{The Low/Hard State}

In this state, the energy spectrum is dominated by a power-law-like
component, with photon index typically $\sim$1.6, showing a clear
high-energy cutoff at approximately 
$\sim$100 keV. An additional weak very
soft component, probably associated with the thermal disc, might be 
observable below 1 keV if the insterstellar absorption is not too high.
This state is positively associated only with the early and final phases
of the outburst of transient systems, which indicates its association to
relatively lower values of mass accretion rate. In the timing domain, 
the PDS shows strong ($\sim$30\%--40\% fractional rms) variability, consisting
of few band-limited components and sometimes a low-frequency QPO, whose
characteristic frequencies appear to be correlated with accretion rate.

The low/hard X-ray state (LS) in canonical X-ray binaries is usually,
probably ubiquitously, associated with flat-spectrum radio emission,
which varies in a correlated way with the X-ray emission (Fender 2001,
2004). Specifically, the radio spectral index $\alpha \geq 0$ (where
$\alpha = \Delta \log S_{\nu}/\Delta \log \nu$, i.e., $S_{\nu}
\propto \nu^{\alpha}$). The radio luminosity varies with the soft
X-ray luminosity approximately as $L_{\rm radio} \propto L_{\rm
X}^{0.7}$ (Corbel et al. 2003, Gallo et al. 2003) until the transition
to softer states, when it drops dramatically (Fender et al. 1999,
Corbel et al. 2001, Gallo et al. 2003). In the bright hard state
system Cygnus X-1, this radio emission has been resolved into a steady
milliarsecond-scale jet (Stirling et al. 2001).

\subsubsection{The High/Soft State}

Here the energy spectrum is dominated by a thermal component, usually
modeled as a disc-blackbody with an inner temperature of 1--2 keV. A
much weaker power-law component is present, with a steep photon
index. No apparent high-energy cutoff is observed in high
signal-to-noise spectra. Very little variability is observed, with the
PDS showing a power-law component with a few percent fractional rms. This
state is usually found in the central parts of outbursts of
transients, indicating its association to a higher mass accretion rate
than the LS.

The high/soft state HS) is associated with a dramatic drop in the radio
emission compared to that observed in the LS, essentially
to undetectable levels (Fender et al. 1999, Gallo et al. 2003),
representing a drop in radio luminosity by at least a factor of 30.
Sometimes optically thin radio emission is associated with the
HS, and is probably associated with shocks in material
distant from the binary core, which is itself no longer a radio source
(e.g., Gallo et al. 2004).

\subsubsection{The Intermediate/Very High State}

The name of this state indicates its dual nature and the complex history
behind it. The energy spectrum is characterized by intermediate properties
between the LS and the HS, with both a strong thermal component and a 
power-law component with $\Gamma\sim$2.5. The relative contribution of
these two components to the 1--10 keV flux can vary between 10\% and 90\%.
As in the HS case, no high-frequency cutoff has been measured in the
hard component (Grove et al. 1998).
The PDS is either a band-limited noise with a strong low-frequency QPO
or a weaker steep power-law component, sometimes with a strong and narrow QPO
at a frequency of $\sim$ 6 Hz (see Nespoli et al. 2003; P. Casella, T. Belloni, J. Homan \& L. Stella,
submitted manuscript; J. Homan \& T. Belloni, manuscript in preparation). This is the
state when high-frequency QPOs have been observed in a few systems.
>From the analysis of the outburst of a number of transients, 
it is emerging that the appearance of this state is both 
related and unrelated to the variation of mass accretion rate. The 
transition between LS and HS and the reverse transitions, happening at
very different accretion rates, take place through a period
of intermediate/very high state (VHS/IS). On the other hand, clear, brief instances of VHS/IS are observed
during the HS periods, indicating that another parameter different from 
$\dot M$ can trigger a transition. Notice that in many systems, for 
instance, XTE J1550-564 (Cui et al. 1999) and GX 339-4 (Belloni et al. 2002), the PDS of the LS at the beginning of the
outburst evolve in a smooth manner, with all characteristic frequencies 
increasing in time, and this smooth evolution continues into the VHS/IS,
despite the fact that the spectral properties change considerably.

The association of radio emission with the VHS/IS is unclear. Corbel
et al. (2001) demonstrated that when the X-ray binary XTE J1550-564
entered a relatively soft IS, the radio emission dropped in a way
reminscent of the radio quenching in the HS. Furthermore, the radio quenching in Cyg X-1 (Gallo et al. 2003)
may also be associated with a transition to the IS and not the
true HS. On the other hand, many transient outbursts are
associated with a rapid rise to the VHS, and such outbursts are nearly
always associated with bright optically thin events (Fender \&
Kuulkers 2001).

\subsection{GRS 1915+105 A/B/C States and Other Systems}

Three states are found in GRS 1915+105 and three main states are known
in other BHCs. It is natural to compare them to see if they are
similar or perhaps even the same states under different names,
although the timescales of transitions are obviously much different.
A comparison based mainly on timing properties was done by Belloni
(1999b) and Reig et al. (2003). Table \ref{pablo_tab} shows the six states in terms of X-ray spectral and timing
properties, with a comment on the jet properties.  Spectrally, it is
clear that the maximum inner temperature of the disc component is
higher in GRS 1915+105, related to the high value of the accrtion
rate. However, the disc component behaves generally in a similar way
during the state C and the LS.  Indeed, there is the suggestion that
the inner radius disk during the LS of BHCs is larger than in the
other states (see, e.g., Esin et al. 1997), although the physical model
associated to this might be different. Because the timing properties of
LS and VHS/IS seem to be linked (see above), it is therefore possible
that the hardest intervals of GRS 1915+105 are instances of a LS, and
indeed the power-law index does reach LS-like values. However, the absence of a high-energy cutoff measured by {\it CGRO}/OSSE would argue
against this interpretation (Zdziarski et al. 2001). In
Table \ref{pablo_tab} we put a question mark in the corresponding box,
as {\it RXTE} measurements shown before seem to indicate the actual presence
of a high-energy cutoff that can reach values as low as a few dozen
keV.  Both states A and B have variability properties that would
associate both of them to the VHS in its steep version (see above).

%Insert Table 2 here

The overall pattern of behavior of the jets in GRS 1915+105 also
seems to be qualitatively very similar to more ``normal'' black hole
X-ray binaries. That is, it follows the basic rules of hard states
associated with radio emission/jet production, soft states
associated with a quenching of the radio emission/cessation of jet
production, and (rapid) transitions between hard and soft states
associated with bright optically thin events. The direct imaging of
the steady jets in plateau states seems to be directly comparable to
the imaging of such jets (albeit much weaker) from Cyg X-1 in the LS.
Finally, the superluminal ejection events are likely to be the same
phenomenon as has been directly imaged during bright transient
outbursts for some other systems (most notably GRO J1655-40, but see
references in Fender \& Kuulkers 2001).

Our conclusion is that probably all three states of GRS 1915+105 are
instances of something similar to the VHS/IS observed in other systems,
associated to the high accretion rate value for this source, although 
during the hardest intervals a LS might be reached.

\begin{table}%
\def~{\hphantom{0}}
\caption{Schematic comparison between the A/B/C states of GRS 1915+105 and
	the canonical states of BHCs. (adapted from 
	Reig et al. 2003)}\label{pablo_tab}
\begin{tabular}{@{}lcccccc@{}}%
\toprule
Parameter& LS & HS & VHS & State A & State B & State C\\
\colrule
kT$^a$ (keV)     & $<$ 1   & $\sim$ 1 & 1--2       & 1.8        & 2.2        & 0.8\\
$\Gamma$$^b$     & 1.5--2  & 2--3     & $\sim$ 2.5 & $\sim$ 3.5 & $\sim$ 3.1 & 1.8--2.5\\
Cutoff$^c$     &  yes    &    no    &     no     &    no     & no         & no?\\
Noise$^d$       &FT             &PL             &FT \& PL       &PL     &PL
&FT\\
$\nu_{\rm break}$&$\le$ 1 Hz    &-              &$\ge$ 1 Hz&1--3 Hz&-   &$\sim$10 Hz \\
rms             &30--50\%       &$<$ 3\%        &1--15\%        &5\%--10\%&5\%--10\%&10\%--20\%\\
QPO             &no             &no             &1--10 Hz       &no     &no
&1--10 Hz \\
%radio           &jet            &no             &bright         &weak   &weak
%&jet \\
\botrule
\multicolumn{7}{l}{$^a$Temperature of the disc blackbody component.} \\
\multicolumn{7}{l}{$^b$Power-law index of the hard spectral component.} \\
\multicolumn{7}{l}{$^c$Presence of a high-energy cutoff.} \\
\multicolumn{7}{l}{$^d$Type of noise in the PDS: FT stands for flat-top and PL
for power-law.}  \\

\end{tabular}
\end{table}

\subsection{Active Galactic Nuclei}

We have discussed the disc-jet coupling in GRS 1915+105
and compared it with other accreting stellar-mass black holes in X-ray
binary systems. But what about the connection to supermassive black
holes (SMBH) in AGN? One of the
great potentials of X-Ray
Binary (XRB) research has been the possibility that the physics can be
scaled up in size and timescale to AGN, and thereby provide us with
insights into the workings of the most powerful engines in the
universe. The discovery of the large power and broad ubiquity
of jets from BHC XRBs has only heightened this interest. With a
general assumption that physical timescales scale with the mass of the
black hole (e.g., Sams, Eckart \& Sunyaev
1996a and references therein), even one hour of
disc-jet coupling in a 10 M$_{\odot}$ black hole probes an equivalent
timescale of $\sim 1000$ years in a $10^8$ M$_{\odot}$ SMBH.

GRS 1915+105 displays a steady jet during plateau states. The X-ray and radio luminosities during this
state are compatible with the universal relation found for the
LS of X-ray binaries, namely $L_{\rm radio} \propto L_{\rm
X}^{0.7}$ (Gallo et al. 2003). Merloni et al. (2003) and Falcke et
al. (2003) have, with the inclusion of a mass term, found that a
closely related expression can fit the 
$L_{\rm radio}$:$L_{\rm
X}$:$M_{\rm BH}$ plane for X-ray binaries (including GRS 1915+105) and
AGN (note that Falcke et al. 2003 restrict themselves to low
luminosity AGN). Therefore, the steady jet in the plateau states is
directly comparable with those in AGN not only qualitatively but
quantitatively (see also earlier discussion in Falcke \& Biermann
1996). Furthermore, Maccarone et al. (2003) present evidence for
quenching of the radio emission in AGN in the same range of
Eddington-fraction X-ray luminosities (approximately 2\%--10\%
Eddington) as observed in the soft states of X-ray binaries (Figure 25).

%!**COMP: Please insert Figure 25 here**!

The timescales associated with, for example, state transitions and transient jet formation
can vary widely (by factors of at least $10^4$) for black holes of
(probably) very similar masses, so its not obvious how to scale up to
SMBH, at least until the coupling with other parameters, most likely
accretion rate, is understood. Nevertheless, Marscher et al. (2002)
have claimed to see an equivalent of the hard dip--radio flare cycle,
made famous by GRS 1915+105, in the AGN 3C 120 (Figure 26, see color insert). They find dips in the X-ray lightcurve, with
a frequency of $\sim 1$/year, with corresponding hardening of the
X-ray spectrum. They further connect these dips to superluminal
ejection events directly imaged with VLBI. The ratio of masses $10^{5}
\leq M_{\rm 3C 120} / M_{\rm GRS 1915+105} \leq 10^{6}$ means that
events of timescales $\sim 1$ year in 3C 120 would correspond to
timescales of seconds or minutes in GRS 1915+105, close to those
observed. Note that the observation of these dips being associated
with highly relativistic ejections is consistent with the conclusion
drawn earlier that the oscillation events are probably associated with
large bulk velocities.

%COMP: Figure 26 is color

We can therefore conclude that the physics of disc-jet coupling in
XRBs and AGN are very closely linked, and that by studying the nearby
XRBs---and GRS 1915+105 in particular---we may understand the flow
of energy to and away from SMBH at the centers of galaxies over
timescales of millions, or even billions, of years.

\section{Interpretation: Many Flavors of Disc-Jet Coupling or Just One?}

On the basis of our detailed studies of GRS 1915+105, we can make some
progress with our understanding of the coupled accretion:outflow
progress in accreting black hole systems. 

\subsection{Hard States and Jet Formation}

As discussed above, protracted spells in hard X-ray states are associated with steady jet
formation and long periods in soft states are not. Such
observations, also in the context of similar behavior in other X-ray
binaries (Fender 2004 and references therein) have been considered as
strong evidence for the association of magnetohydrodynamic (MHD)
outflows with large scale-height accretion flows (e.g., Meier 2001,
Meier et al. 2001). The hard X-ray component in such states is
generally supposed to arise in some form of Comptonization
(e.g., Sunyaev \& Titarchuk 1980, McClintock \& Remillard 2004), in
which case a clear connection between the ``base'' of the outflow and
this Comptonizing corona is required (see discussions in Muno et
al. 2001, Zdziarski et al. 2003, Fender 2004). Alternatively, the hard
X-ray emission may arise directly in the jet via Comptonization (see
discussion in Atoyan \& Aharonian 1999) or, more radically, optically
thin synchrotron emission (Markoff et al. 2001, Vadawale et
al. 2001). In either case, a clear coupling between the hot plasma
responsible for the hard X-ray emission, including its associated
strong X-ray variability, and the production of a jet requires
explanation in future models. However, this is only part of the story.

\subsection{On the Trigger of Optically Thin Events}

Empirically, we can also suggest that hard-to-soft state transitions
are associated with resolvable ejection events, but is there a unique
signature in the X-ray emission that tells us precisely
when this event occurs?

Mirabel et al. (1998) suggested that the brief spike seen
toward the end of the state-C dip during the oscillations of
GRS 1915+105 corresponded to the ``launch'' of the radio oscillation
event. During this spike, the X-ray spectrum softened dramatically,
indicating a rapid change in the state of the accretion flow. However,
an overriding problem with seeing clearly the disc-jet coupling in GRS
1915+105 (despite the many advantages the source offers) has been that
the rise, decay, and recurrence times of the synchrotron events are
all comparable.  In general, a
synchrotron event begins to rise before the previous event is fully
decayed, so observing the beginning of the ``flare'' phase is
difficult.

By drawing analogies with other X-ray
binaries whose temporal evolution is a little more
subdued, we can learn more. Fender, Belloni \& Gallo (2004b) have
studied in detail the light curves of GRS 1915+105 and three other
black hole transients---GX 339-4, XTE J1859+226, and XTE J1550-564---to determine at what point the optically thin radio flare
occurred with respect to the X-ray state. The results indicate that
the radio event occurs during a local peak in the X-ray light curve,
which corresponds to the end of a phase of softening in the X-ray
spectrum. In terms of GRS 1915+105, the source makes a transition from
state C to states A/B. In terms of the canonical states of other X-ray
binaries, such as those listed above, the X-ray peak corresponds to
neither the canonical LS or the canonical HS, but to a transition from
a ''soft'' VHS/IS to a ''hard'' VHS/IS (often en route between the LS and
HS).

In many, possibly most, of the more conventional transients, the
spectral evolution of the source around the time of the ejection can
be tracked more clearly than in GRS 1915+105. What is generally
observed is a peak in the LS (e.g., Brocksopp et al. 2002, Maccarone \&
Coppi 2003, Yu et al. 2003).  At this point the strongest
self-absorbed jet is probably being produced. Subsequently, most
sources transit to a softer X-ray peak, which corresponds to the
VHS/IS, and then later to the canonical HS. It is at the VHS/IS peak
that the optically thin radio event occurs; subsequently, the jet is
''off'' until the return to a harder X-ray state. Applying the knowledge
we have gained from these other transients to GRS 1915+105, we assert
that the optically thin ejection event occurs at the state C
$\rightarrow$ state A transition.

\subsection{Jet Power and Velocity}

The power associated with the strongest steady jets (those around the
LS peak in the X-ray light curve) and the subsequent optically
thin events seems to be remarkably close. We can demonstrate this in
the following. Fender, Gallo \& Jonker (2003) have presented a simple
formula for estimating the power in a jet in the LS of an X-ray binary:

\[
L_{\rm J} = A L_{\rm X}^{0.5},
\]

\noindent where $L_{\rm J}$ and $L_{\rm X}$ are the powers in
Eddington units in the jet/outflow and radiation from the
accretion flow (primarily X-rays), respectively. The value of the
constant is estimated as $A \geq 6 \times 10^{-3}$. Note that Malzac, Merloni \& Fabian (2004) have suggested
that all low-state sources are jet-dominated, in which the value of
the normalization is much larger: $A \geq 0.1$.   

Fender, Belloni \& Gallo (2004b) have further compiled estimates of the
power associated with optically thin ejection events as a function of
the corresponding peak (VHS/IS) X-ray luminosity and found a function
of the form

\[
L_{\rm jet} = B L_{\rm X}^{0.4 \pm 0.1},
\]

where the exponent $(0.4 \pm 0.1)$ results from a fit to multiple
independent estimates. These power estimates are made assuming
equipartition on the basis of observations of rise times and luminosities of
radio events (see Fender et al. 2004b for more details). This is in
contrast to the estimates for the steady jets, which are based on
broadband spectra and estimates of radiative efficiency (Fender et
al. 2003 and references therein) or models for the production of
correlated optical and X-ray variability (Malzac et al. 2004).

The normalization for the transients jets, $B \sim 0.5$ (with a large
uncertainty, probably as much as an order of magnitude larger or
smaller), is directly comparable to the normalization $A$ estimated
for the steady jets above (assuming no significant advection of
accretion energy in either case).  Given the completely independent
methods used to estimate the jet power in the two regimes, it is
remarkable that the exponents of the two functions are the same within
uncertainties. In the event that the estimate for $A$ of Malzac et
al. (2004) is closer to reality than the lower limit given in Fender
et al. (2003), then the normalizations are also very close. From this
we draw the conclusion that the power associated with the strongest
self-absorbed jet, and that associated with the subsequent optically
thin flare, are comparable (within an order of magnitude or so). This
strongly suggests a rather smooth function in the relation between jet
power and X-ray luminosity, and not two completely independent
mechanisms for powering the two types of jet.

As noted in Gallo et al. (2003), and hinted at for GRS 1915+105 (see
above), it seems that the velocity of the steady jets is also lower
than that associated with the optically thin ejection events. This is
based on the fact that the spread around a common $L_{\rm jet}
\propto L_{\rm X}^{0.7}$ relation for all black hole XRBs in the LS
requires $v_{\rm LS} \leq 0.8c$ (for a random distribution of viewing
angles), otherwise Doppler boosting would result in a larger
scatter. Fender et al. (2004b) have refined this fit and compiled
multiple pieces of evidence that the Lorentz factor associated with
the optically thin ejections of transient XRBs is $\Gamma \geq
2$. Therefore, although we do not know the form of the function
(e.g., it could be step-like or continuous), it seems clear that the
transient jets have larger bulk Lorentz factors than the steady jets.

\subsection{A Unified (Toy) Model of Disc-Jet Coupling}

In the following, we attempt to construct a semi-quantitative
model for the radio emission in GRS 1915+105 that is hopefully
relevant to other X-ray binaries and to AGN. In order to lay the
foundations for this model, we summarize some of the key physical
indicators:

\begin{itemize}
\item{Prolonged ``hard'' X-ray states (e.g., canonical LS, GRS 1915+105
  state C) are associated with steady
  production of self-absorbed jets.}
\item{Prolonged ``soft'' X-ray states (e.g., canonical HS, GRS 1915+105
  states A and B) are associated with a suppression of
  jet production.}
\item{During the transition between hard and soft X-ray states, an
  optically thin ejection event occurs. In the canonical definitions
  of states, this event occurs at the VHS/IS peak, close to the end of
  a phase of X-ray spectral softening. This corresponds to a
  transition from state C to state A in GRS 1915+105 (the C to B
  channel is not observed; see Figure 8).}
\item{The transient jets associated with these events have
  significantly larger bulk Lorentz factors than those associated with
  the steady jets.}
\end{itemize}

Based on these observational results, we have arrived at the
following semi-quantitative picture of the evolution of the disc-jet
coupling during X-ray binary outbursts in general, and those of GRS
1915+105 in particular. The picture is summarized in Figure 27 (see color insert), and in
the text below.

%COMP: Figure 27 is color

The upper panel in Figure 27 is a hardness-intensity diagram for the
X-ray emission from a given source. The canonical LS and HS correspond
to nearly vertical strips at the extreme right (hard) and left (soft)
of the panel, respectively. Practically the entire range in hardness
between these two canonical states represents VHS/IS of varying
hardness. We suggest that above a certain hardness (phases i and ii) a
jet is produced whose strength correlates with that of the X-rays in
the nonlinear way found by Corbel et al. (2003) and Gallo, Fender \&
Pooley (2003). Below that hardness (phase iv) a jet is not produced;
these two regimes correspond to the regions right and left of the
vertical line in Figure 27, respectively.  We propose that as a source
spectrum softens and it approaches the line from the right, the
velocity of the jet increases monotonically. At first, this increase is
insignificant for the energetics of the outflow, which is not highly
relativistic (bulk Lorentz factor $\Gamma \sim 1$).  However, around
the time that the source crosses the line (phase iii), which generally
corresponds to a peak in the soft X-ray flux, the jet velocity
increases very rapidly ($\Gamma \geq 2$) before the jet is shut
off. Such a rapid increase in jet velocity would produce a shock in
the preexisting, slower-moving, jet, which would manifest itself in
an optically thin flare with large bulk motions.

For most X-ray binaries, an outburst follows a path similar to that
indicated by the solid arrows in Figure 27. The ``quiescent'' state of such
sources can be found by extending the LS branch vertically
downward. Often, sources will exhibit a few secondary flaring states
in the same outburst, associated with further optically thin radio
flares (e.g., GRO J1655-40, XTE J1550-564, XTE J1859+226---see,
e.g., Brocksopp et al. 2001, Homan et al. 2001), but these are usually
of declining strength. It is interesting to note that in the model of
Vadawale et al. (2003), this declining sequence may be naturally
explained by the first event having considerably more material in
front of it with which to interact (because there was a long phase of
steady jet production in the ``quiescent'' state and LS).  In any case,
GRS 1915+105 seems to remain, for most of the time, very close to the
dividing line between jet-producing states and states without
jets. The radio--mm--IR--X-ray ``oscillation'' events (e.g., Figures 3, 13,
21, and 23) for which GRS 1915+105 is so famous may be compared to the
encircled region in the upper-left of the diagram, with the direction
of a typical oscillation being indicated by the dahsed arrow. In this
picture, GRS 1915+105 is repeatedly entering the jet-producing state
then crossing the dividing line to jet-free states. Each transition
from right to left is associated with an optically thin event; those
from left to right are not. Sometimes GRS 1915+105 remains in the hard
state (state C) for considerably longer and may extend across the
upper branch toward the canonical LS. However, the transition from
this state inevitably involves crossing the line from right to left
once more, producing a major flare event.

This model should be considered as a synthesis of previous
observational results and theoretical interpretations (although any
errors resulting from the synthesis are entirely ours!). Mirabel et
al. (1998) first suggested the idea that the X-ray ``spike'' at the end
of hard ``dips'' might correspond to the moment of jet production. We
have expanded on this to show that in other X-ray binaries it is the
(secondary) VHS/IS peak and not the (initial) LS peak that corresponds
to the optically thin radio flare. Meier (1999) discussed a
dramtically (discontinuously) varying jet speed in response to varying
conditions in the accretion flow. 
The internal-shock model based
on varying flow speeds in a semicontinuous jet was originally
developed for the jets of the AGN M87 (Rees 1978), and later
considered for gamma-ray bursts (Rees \& Meszaros 1994) and
generically for radio-loud quasars (Spada et al. 2001). Kaiser et
al. (2000) considered internal shock models for GRS 1915+105 and
Vadawale et al. (2003) explicitly considered ``post-plateau flares'' as
resulting from shocks generated by fast moving ejecta interacting with
a slower-moving outflow produced during the plateau. Esimates of the
jet power in steady and transient phases are more recent (Corbel et
al. 2003; Gallo et al. 2003; Fender et al. 2003, 2004b), as is the
certainty that the transient jets are considerably more relativistic
than the steady jets (Gallo et al. 2003, Fender et al. 2004b).

The model as it stands is mainly phenomenological, but there are some
clues to the underlying physics. For example, the jet velocity seems
to increase as the optically thin accretion disc moves ever closer to
the central accreting object, only becoming significantly relativistic
in the very final stages. This may naturally reflect the escape
velocity of matter from the inner edge of the accretion disc, which
will remain almost constant until it enters regions of significant
spacetime curvature within a few gravitational radii of the black hole
(but see Fender et al. 2004a, where a highly relativistic jet from a
neutron star is in contradiction with the ``escape velocity''
scenario). This is consistent with the evolution of the X-ray colors,
but what exactly happens at this point? It could be that the soft
state, for some reason (e.g., no significant vertical magnetic field or
extreme Compton cooling of relativistic electrons), inhibits jet
formation, and that the optically thin flare was a last gasp from the
jet before it is suppressed. On the other hand, it may be that the
sudden increase in jet velocity/power is enough to force the
accretion flow to settle into a different state, in which case the
state change is a result of, not a cause of, the radio ejection
event. The similarity of the formulae for the power associated with
the two types of jet may argue for a common underlying mechanism,
which is simply the release of gravitational potential in the
accretion flow. Alternatively, the boost in the jet velocity, and
possibly power, for the very smallest disc radii may be in response to
a sudden additional contribution from the black hole spin.

Finally, what makes GRS 1915+105 so special? As noted above, the
evolution of outbursts from other X-ray transients typically follow
the path of the solid lines in Figure 27. Although some (see, e.g., Homan et
al. 2001, Brocksopp et al. 2002) may make a small number of repeated
phase iii transitions, resulting in several ejection events, they
have all returned to relative obscurity within a year or so. GRS
1915+105 has, on the contrary, been very luminous and active for $\geq
10$ years. It seems clear that its unusual behavior is a result of
a very high (at times probably super-Eddington) accretion rate being
maintained for a very long time, probably as a consequence of binary
evolution (the orbital period of GRS 1915+105 is exceptionally long for
an X-ray transient). In this sense, it is comparable to the brightest Quasars
among the sample of AGN.

\section{Conclusions}

In this review, we discuss at length the
observational properties of GRS 1915+105 in the X-ray,
IR, and radio bands one decade after the
discovery of highly relativistic jets from the source, the first in
our Galaxy.  A connection between the X-rays and the
longer-wavelength (IR and radio) emission in this
source is unambiguous, and as a result it has become one of the key
systems for our understanding of the disc-jet coupling. Placing the
system into context, we have argued that its properties are not that
dissimilar from the X-ray states observed in other black hole
binaries. Beyond this, the relation of disc-jet coupling in X-ray
binaries to that in AGN is now finally on
a quantitative footing, and we can be confident that insights drawn
from GRS 1915+105 will be relevant for the disc-jet connection around
black holes of all masses.  Consequently we have proposed a toy model
for the disc-jet coupling in GRS 1915+105---heavily inspired by other,
earlier works---that requires a minimum (i.e., one)
different mode of jet formation and, we hope, is consistent
with the picture for other black hole systems.

\section*{Acknowledgements}

The authors thank Sergio Campana, Elena Gallo,
Marc Klein-Wolt, Thomas Maccarone, and Simone Migliari for many useful
comments on earlier drafts of this review.

%  ================================================================

%Figure 1
\begin{figure}
\centerline{\psfig{figure=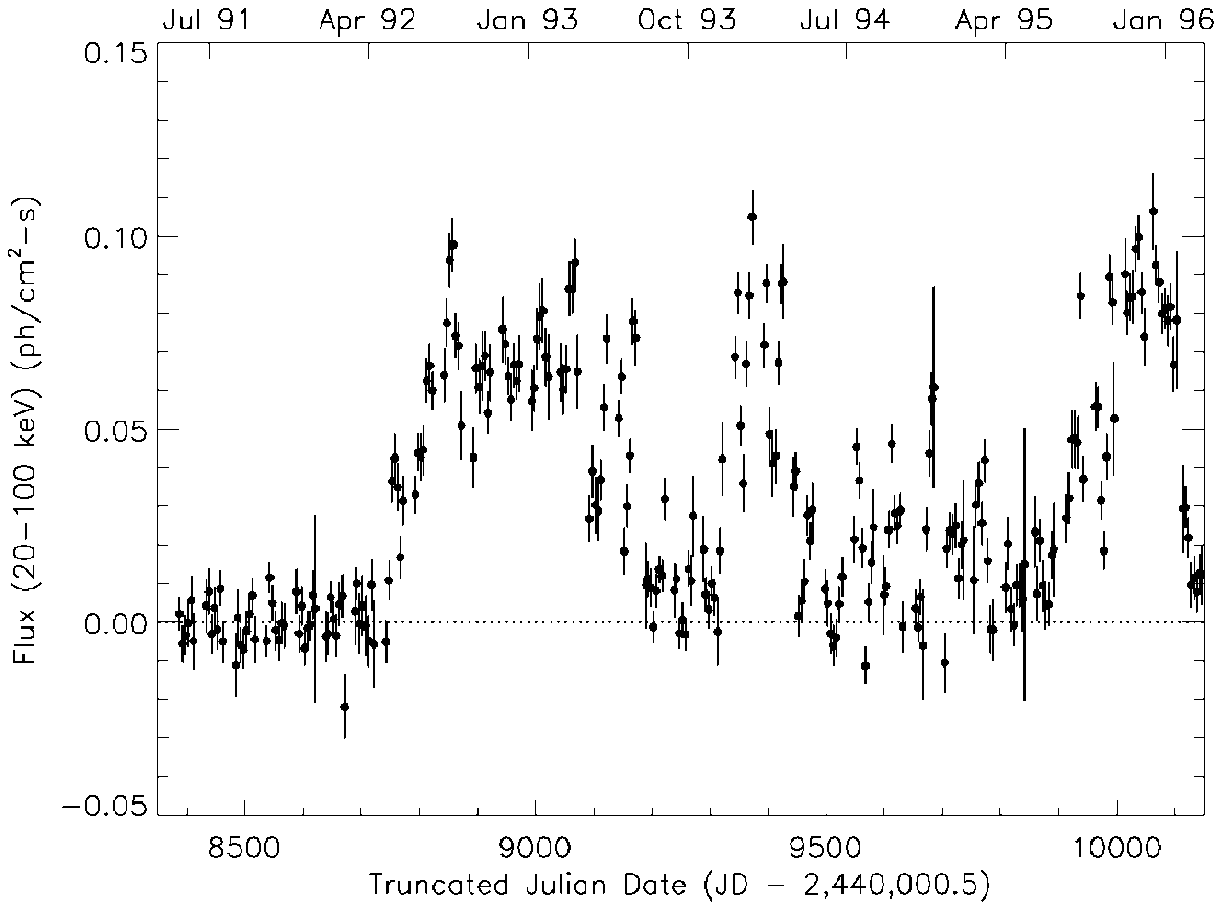,width=9cm}}
\caption{BATSE 20--100 keV light curve of GRS 1915+105 in 5-day bins. Flux
conversion was done with a $\Gamma$=2.5 power-law model
(from Harmon et al. 1997). Time ranges from mid-1991 to the launch of
{\it RXTE}.}
\label{batse}
\end{figure}

%Figure 2
\begin{figure}
\centerline{\psfig{figure=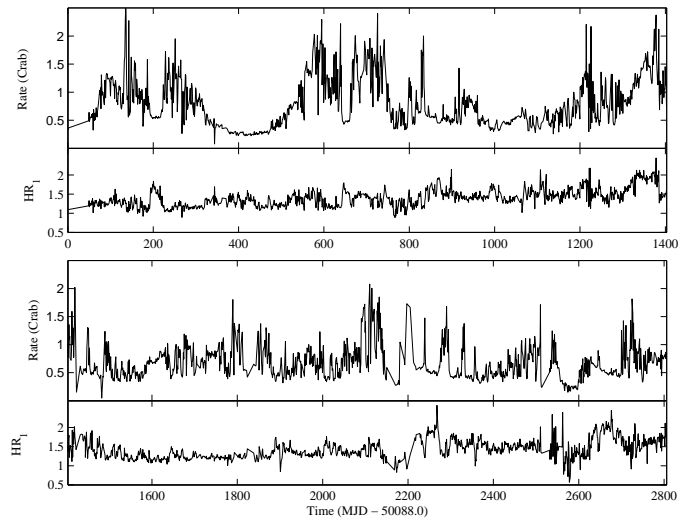,width=9cm}}
\caption{{\it RXTE}/ASM light curve, in 1-day bins, from the start of the
{\it RXTE} mission up to October 12, 2003.}
\label{asm}
\end{figure}

%Figure 3
\begin{figure}
\centerline{\psfig{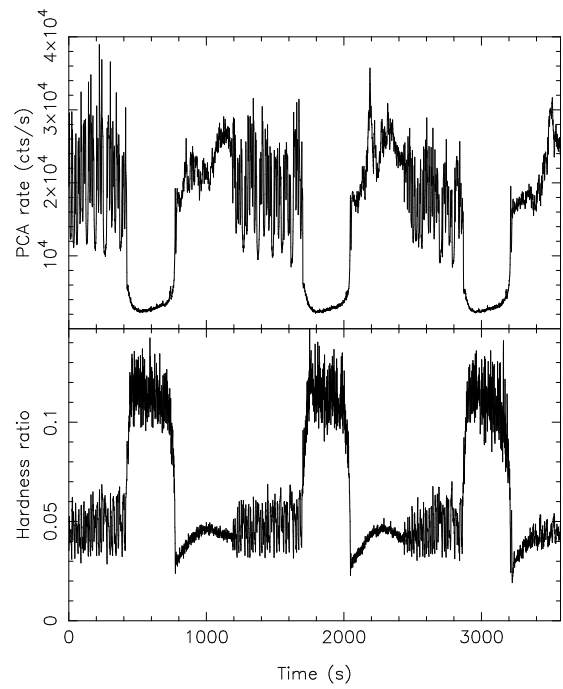}}
\caption{
{\it RXTE}/PCA light curve of GRS 1915+105 from October 7, 1996, with
corresponding hardness ratio (13.3--60.0 keV/2.0--13.3 keV) curve
(from Belloni et al. 1997a).}
\label{bell97a}
\end{figure}

%Figure 4
\begin{figure}
\centerline{\psfig{figure=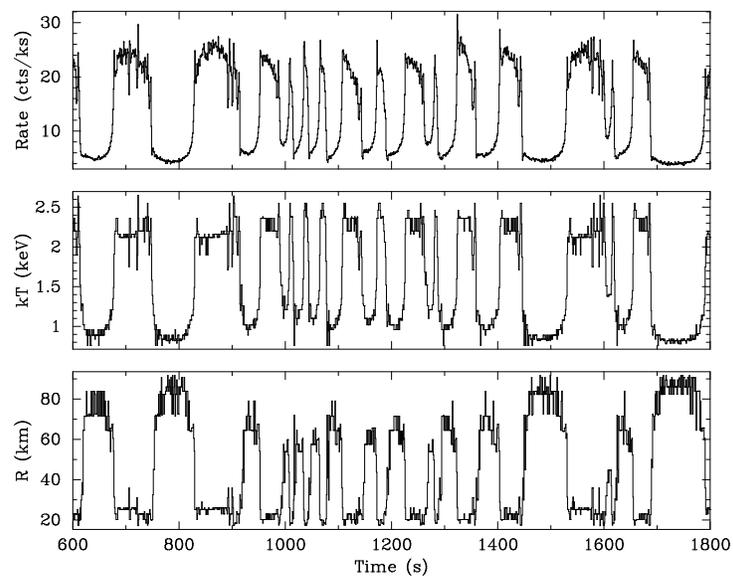,width=9cm, angle=270}}
\caption{Upper panel:
{\it RXTE}/PCA light curve of GRS 1915+105 from June 18, 1997.
Middle and lower panels: corresponding inner radius and temperature
(from Belloni et al. 1997b).}
\label{bell97b1}
\end{figure}

%Figure 5
\begin{figure}
\centerline{\psfig{figure=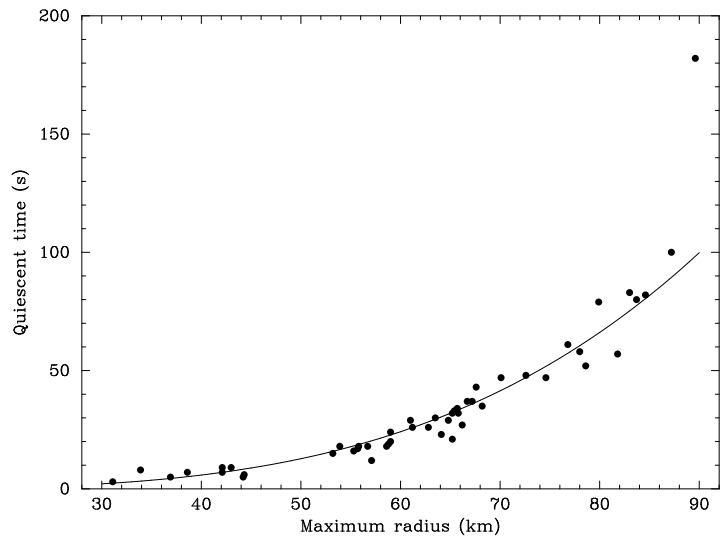,width=9cm, angle=270}}
\caption{Correlation between total length of a hard event and maximum inner radius of
the disc from the data shown in Figure \ref{bell97b1}. The line is the best
fit with a power-law with fixed index 3.5
(from Belloni et al. 1997b).}
\label{bell97b4}
\end{figure}

%Figure 6
\begin{figure}
\centerline{\psfig{figure=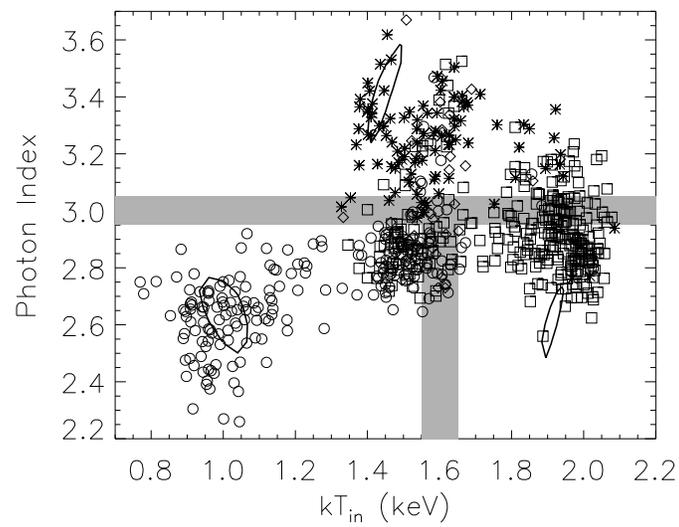,width=9cm}}
\caption{GRS 1915+105 in the kT$_{in}$-$\Gamma$ plane for the {\it RXTE}/PCA observation of September 9, 1997  (from Markwardt, Swank \& Taam 1997). Different symbols correspond to different timing behavior. The gray bands divide the plot into three sections, corresponding to the three states defined by the authors.}
\label{markwardt_states}
\end{figure}

%Figure 7
\begin{figure}
\centerline{\psfig{figure=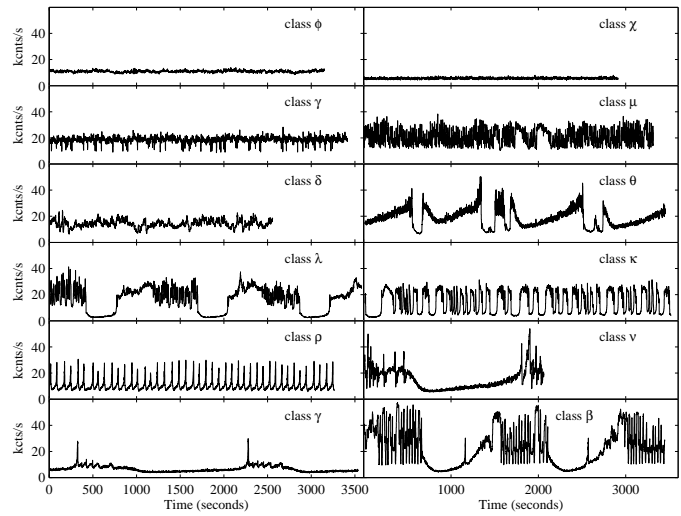,width=9cm}}
\caption{Sample {\it RXTE}/PCA 1-s light curves for six of the 12 classes defined in Belloni et al. (2000).}
\label{class}
\end{figure}

%Figure 8
\begin{figure}
\centerline{\psfig{figure=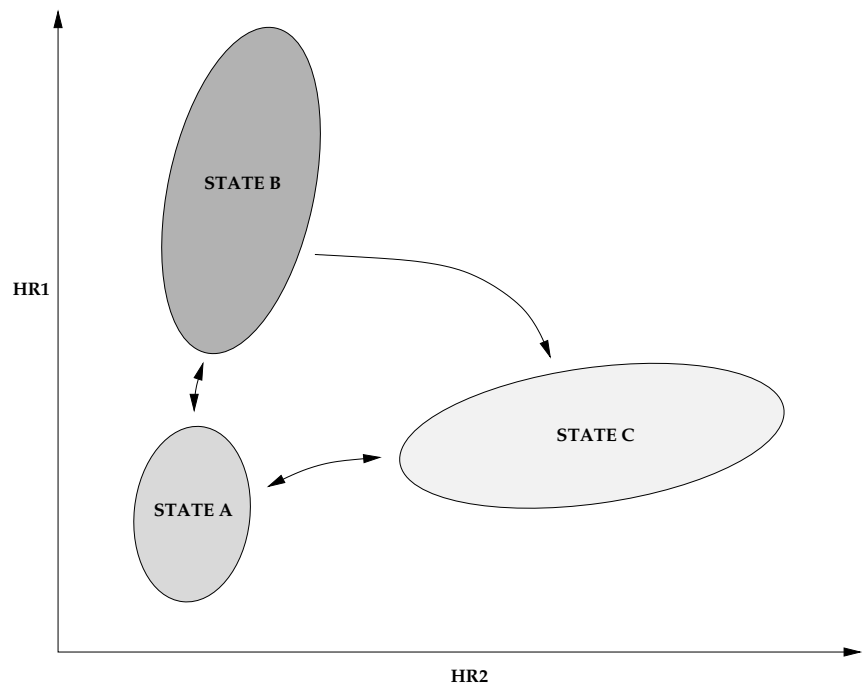,width=9cm, angle=270}}
\caption{Schematic PCA color-color diagram showing the basic A/B/C 
states and their observed transitions. 
The X-ray colors are defined in the following way. HR1 is the ratio
of the counts in the 5--13 keV band over those in the 2--5 keV band. HR2
is the ratio 13--60 keV over 2--5 keV.  From Belloni et al. (2000).}
\label{abc}
\end{figure}

%Figure 9
\begin{figure}
\centerline{\psfig{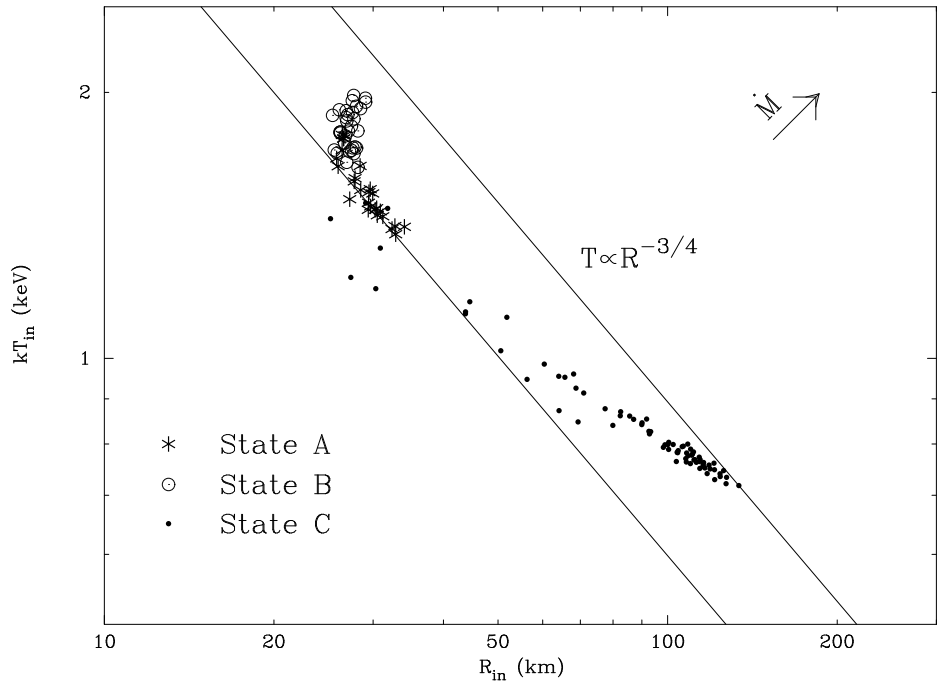}}
\caption{Inner disc temperature versus inner disc radius for two intervals of
	instability of a representative
	class-$\beta$ observation of GRS 1915+105. The lines indicate two
	levels of constant accretion rate (increasing in the direction of
	the arrow (from Migliari \& Belloni 2003).}
\label{migliari}
\end{figure}

%Figure 10
\begin{figure}
\centerline{\psfig{figure=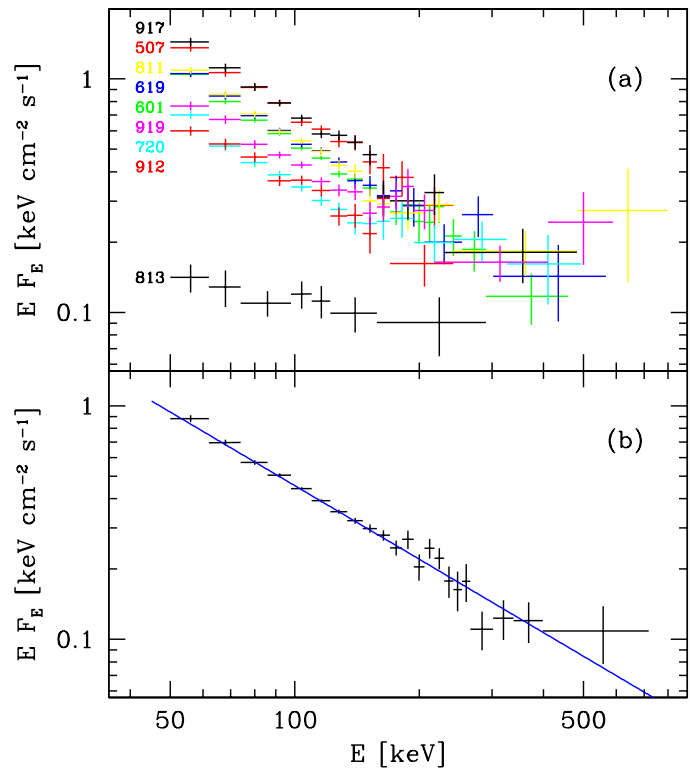,width=9cm}}
\caption{({\it a}) OSSE spectra of GRS 1915+105 deconvolved with a power-law model.
	The spectrum labeled 813 corresponds to the probable B state
	observation. ({\it b}) Average of the above with power-law best fit.
	From Zdziarski et al. (2001).}
\label{zdz}
\end{figure}

%Figure 11
\begin{figure}
\centerline{\psfig{figure=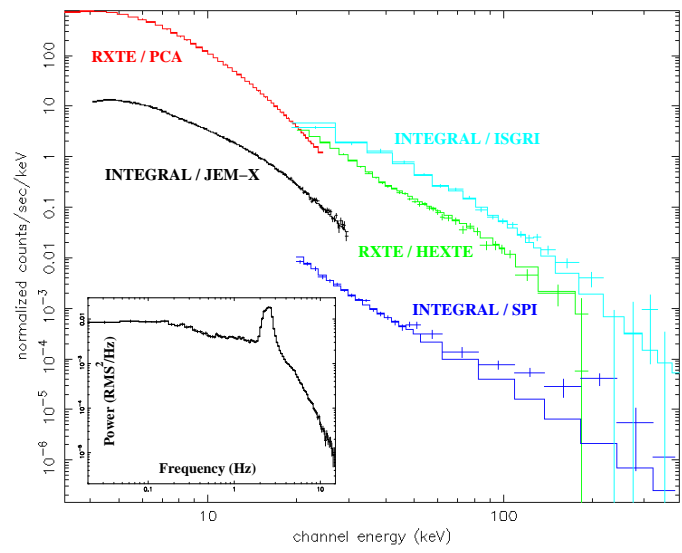,width=9cm}}
\caption{Combined {\it RXTE} PCA+HEXTE and INTEGRAL ISGRI+SPI spectra of 
	GRS 1915+105 during a plateau period on April 2, 2003.
	The inset shows a PCA power density spectrum typical of
	C state.}
\label{integral}
\end{figure}

%Figure 12
\begin{figure}
\centerline{\psfig{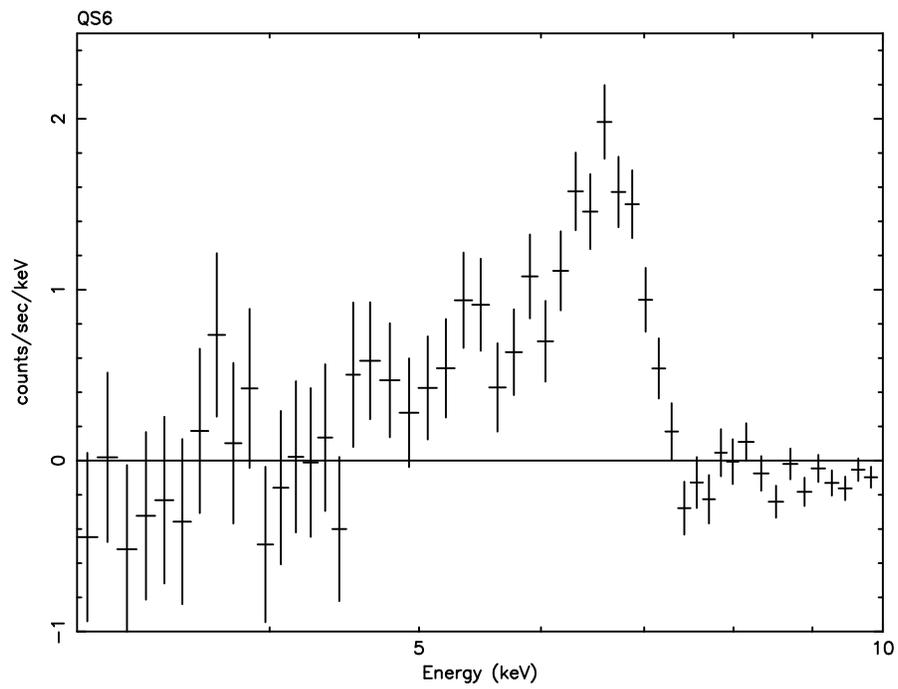}}
\caption{Residuals for one of the BeppoSAX observation intervals showing
	the presence of a broad skewed emission line (from
	Martocchia et al. 2002).}
\label{martocchia}
\end{figure}

%Figure 13
\begin{figure}
\centerline{\psfig{figure=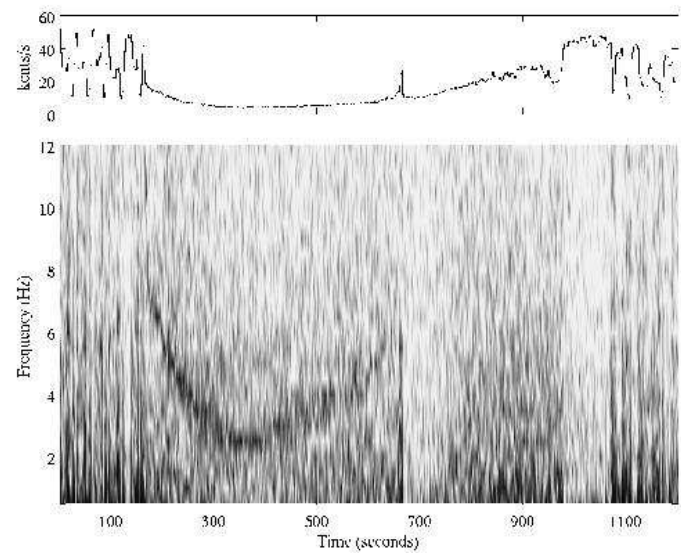,width=9cm}}
\caption{Top panel: PCA light curve from October 31, 1997 (class $\beta$) with
	2-s bin size. 
	Bottom panel: corresponding spectrogram (in logarithmic gray scale).
	Black corresponds to high power values.}
\label{spectrogram}
\end{figure}

%Figure 14
\begin{figure}
\centerline{\psfig{figure=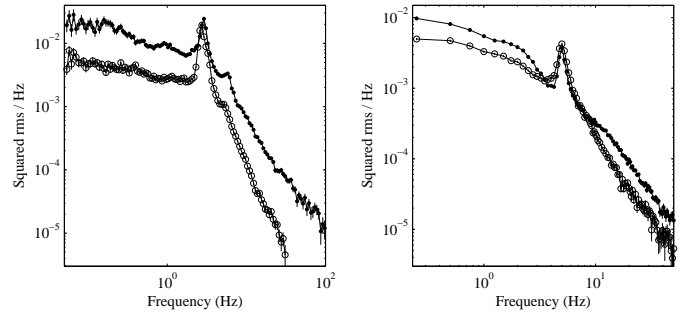,width=9cm}}
\caption{Left panel: PDS of two PCA observations from plateaux periods. Filled
	circles are from August 14, 2000; open circles from 
	July 23, 1996 (adapted from Trudolyubov 2001). Right panel:
	two examples at a higher QPO frequency ({\it open circles}: August 18, 1996; 
	{\it filled circles}: October 23, 1996).}
\label{trudo}
\end{figure}

%Figure 15
\begin{figure}
\centerline{\psfig{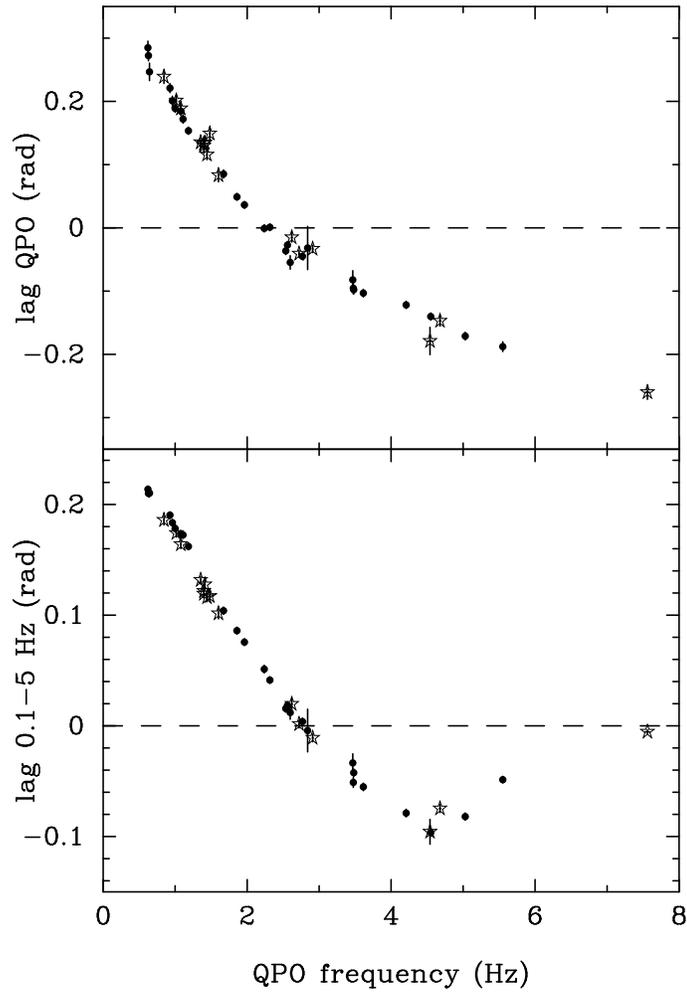}}
\caption{Phase lags associated to a sample of radio-loud 
	plateaux observations of GRS 1915+105 (from Reig et al. 2000).
	Top panel: phase lags at the QPO frequency. Bottom panel:
	phase lags of the continuum between 0.1 and 5 Hz. Positive lags
	correspond to the hard flux lagging the soft flux.}
\label{reig_lags}
\end{figure}

%Figure 16
\begin{figure}
\centerline{\psfig{figure=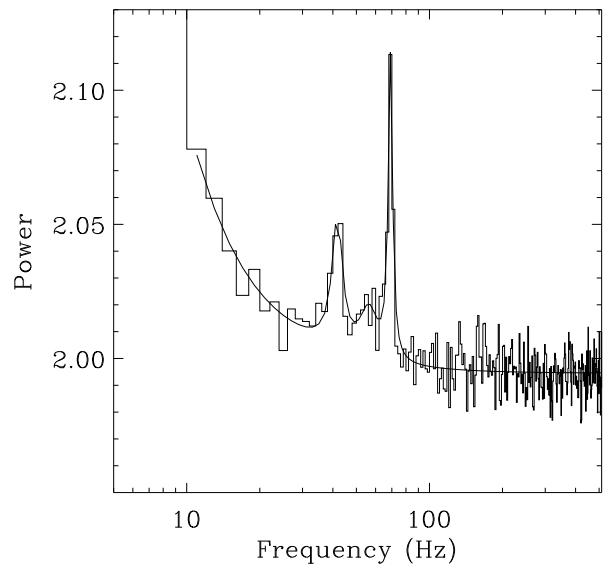,width=9cm}}
\caption{Average power density spectrum of the five PCA observations in
        1997 when high-frequency QPOs were detected. Both the 40 Hz and
        the 69 Hz peaks are visible (from Strohmayer 2001).}
\label{tod}
\end{figure}

%Figure 17
\begin{figure}
\centerline{\psfig{figure=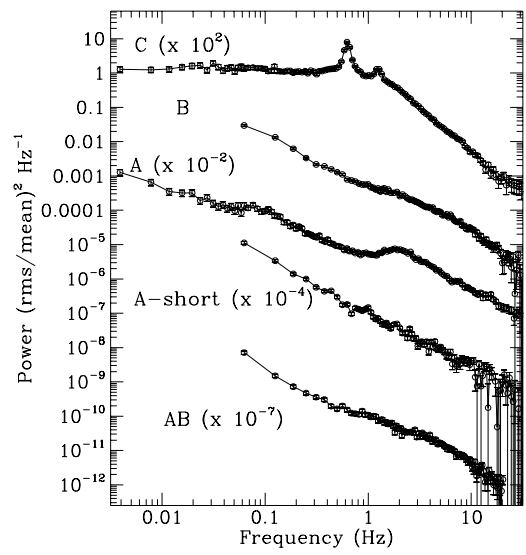,width=9cm}}
\caption{Representative PDS for the three states of GRS 1915+105. A-short refers to short intervals of state A. AB is from a smooth A--B  transition during class $\beta$ (from Reig et al. 2003).}
\label{reig_pds}
\end{figure}

%Figure 18
\begin{figure}
\centerline{\psfig{figure=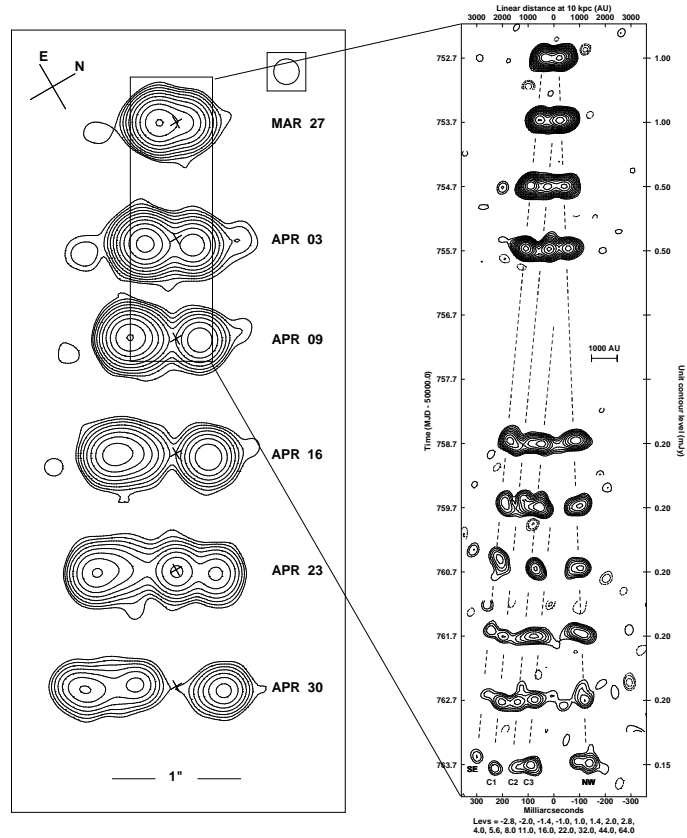,width=12cm,angle=270}}
\caption{Major (superluminal) relativistic ejections from GRS
  1915+105. The left panel shows VLA observations over a period of
  $\sim$ one month in 1994, from Mirabel \& Rodr\'iguez (1994). The right panel
  shows higher-resolution MERLIN imaging over a period of $\sim 12$
  days in 1997, from Fender et al. (1999a).}
\label{rel_jets}
\end{figure}

%Figure 19
\begin{figure}
\centerline{\psfig{figure=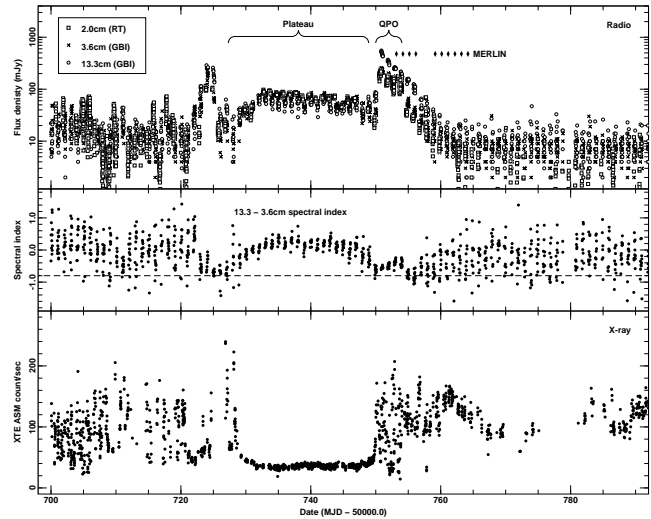,width=9cm}}
\caption{A plateau state in GRS 1915+105, from Fender et al. (1999). The top panel shows radio monitoring at three frequencies, from the Green Bank Interferometer (GBI) and Ryle Telescope (RT). The plateau
phase lasts from approximately MJD 50730 to MJD 50750, and is associated with a flat radio spectrum ({\it middle panel}). The plateau is
associated with steady X-ray emission ({\it lower panel}), which has a hard
spectrum (state C).  Prior to the plateau is a small medium strength
optically thin flare (preplateau flare in Klein-Wolt et al. 2002), and following the plateau is a larger optically thin flare (postplateau flare), which in this case was directly
resolved into a relativistic ejection event (tickmarks indicate the epochs of the MERLIN observations presented in the of Figure 18). Following the first postplateau flare was a period of
four days of X-ray dip/radio oscillation cycles, indicated as QPO. Other occasions of plateaux have resolved the flat-spectrum radio emission into a steady jet such as those presented in Figure {\ref{vlba}}.}
\label{plateau}
\end{figure}

%Figure 20
\begin{figure}
\centerline{\psfig{figure=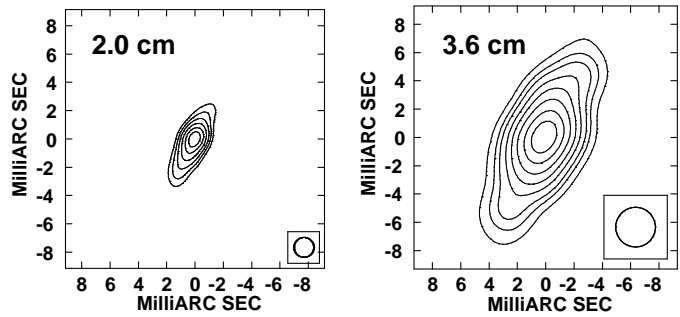,width=9cm}}
\caption{Compact core jet associated with the plateau---steady,
  hard X-ray and radio emission---state in GRS 1915+105 in April 2003
  (adapted from Fuchs et al. 2003b). A similar jet in a plateau state
  in 1998 was reported by Dhawan et al. (2000).}
\label{vlba}
\end{figure}

%Figure 21
\begin{figure}
\centerline{\psfig{figure=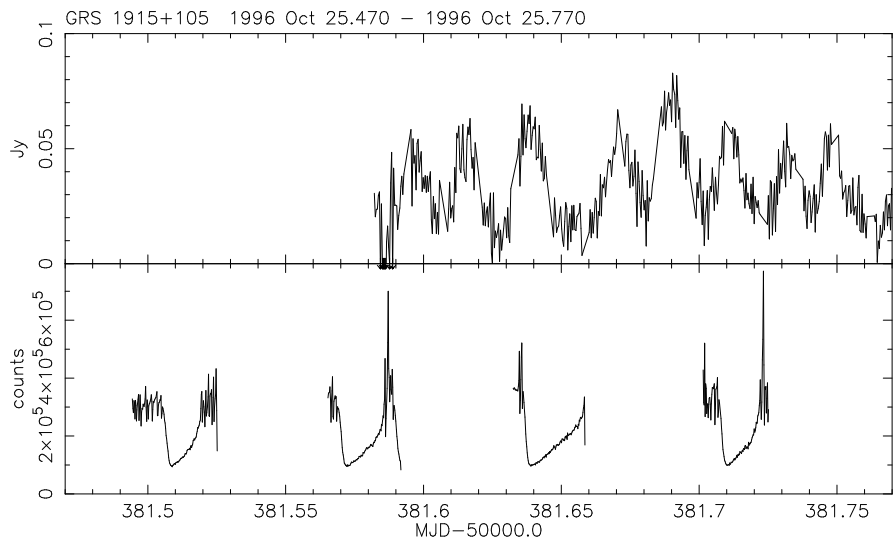,width=7cm, angle=270}}
\caption{Simultaneous radio and X-ray observations of GRS 1915+105 in
  1996, from Pooley \& Fender (1997). Despite the patchy X-ray
  coverage there is a clear hint of an association between the
  semiregular X-ray dipping behavior and the radio oscillations.}
\label{}
\end{figure}

%Figure 22
\begin{figure}
\centerline{\psfig{figure=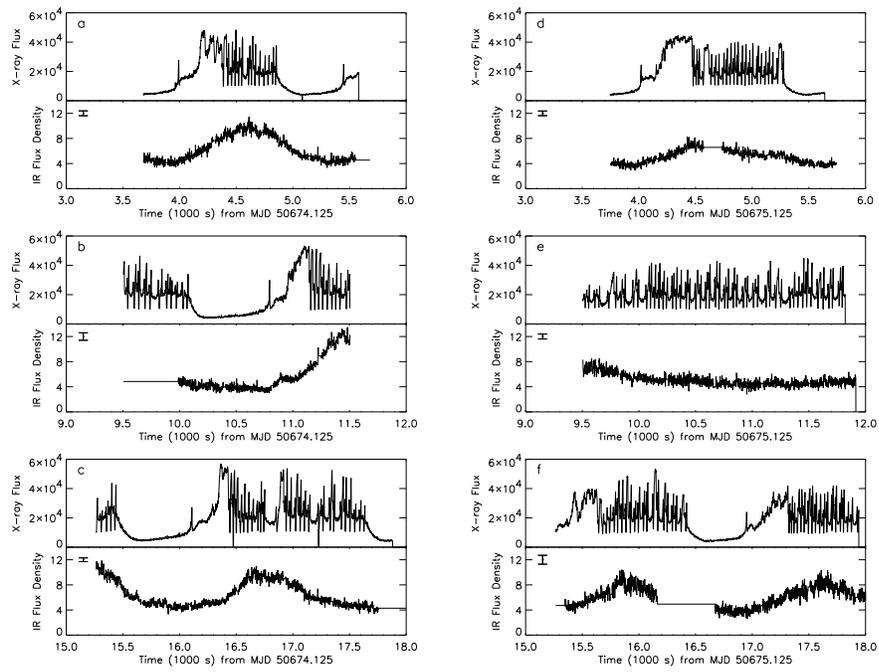,width=9cm,angle=90}}
\caption{Simultaneous X-ray and IR observations of GRS 1915+105
  in August 1997, from Eikenberry et al. (1998a).}
\label{eik}
\end{figure}

%Figure 23
\begin{figure}
\centerline{\psfig{figure=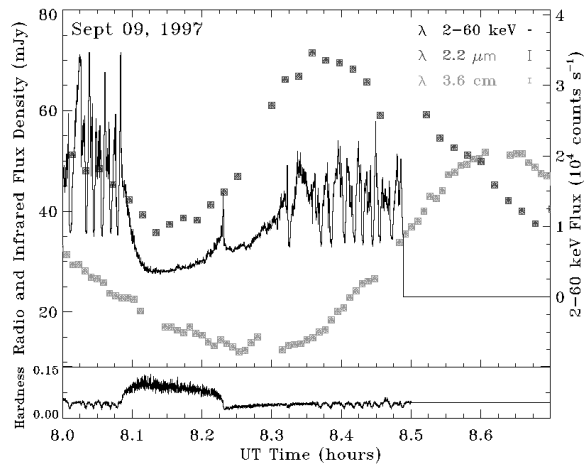,width=9cm}}
\caption{X-ray/IR/radio light curve of GRS 1915+105 during a
  hard X-ray dip, from Mirabel et al. (1998). The radio event is
  certainly associated with the cycle, and the IR precedes it by
  $\sim 20$ min---probably indicating that the IR is
  synchrotron emission from much closer to the base of the jet than
  the radio. Note the spike in the X-ray light curve at which point
  the X-ray hardness drops dramatically, and has been suggested by
  Mirabel et al. (1998) as the time of launch of the relativistic jet.}
\label{m98}
\end{figure}

%Figure 24
\begin{figure}
\centerline{\psfig{figure=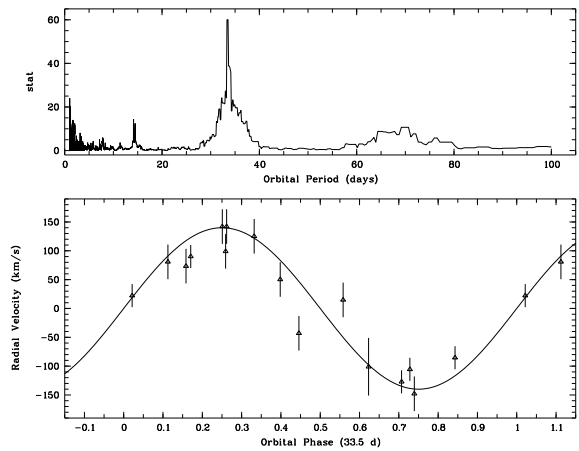, angle=270,width=9cm}}
\caption{Orbital parameters of GRS 1915+105 from IR spectroscopy
(Greiner et al. 2001). The top panel shows the most likely orbital
  period, of 33.5 days, for the binary. The lower panel shows the
  radial velocity data folded on this orbital period. From these data
  a mass function---corresponding to an absolute lower limit on the
  mass of the accreting object---of $9.5 \pm 3.0$ M$_{\odot}$ can be
  derived. Assuming a companion mass of $\sim 1.2$ M$_{\odot}$ and an
  orbital inclination equal to the angle the jets make with the line
  of sight, i.e., $\sim 70^{\circ}$, then Greiner et al. estimate a
  mass of $14 \pm 4$ M$_{\odot}$ for the black hole.}
\label{radvel}
\end{figure}

%Figure 25
\begin{figure}
\centerline{\psfig{figure=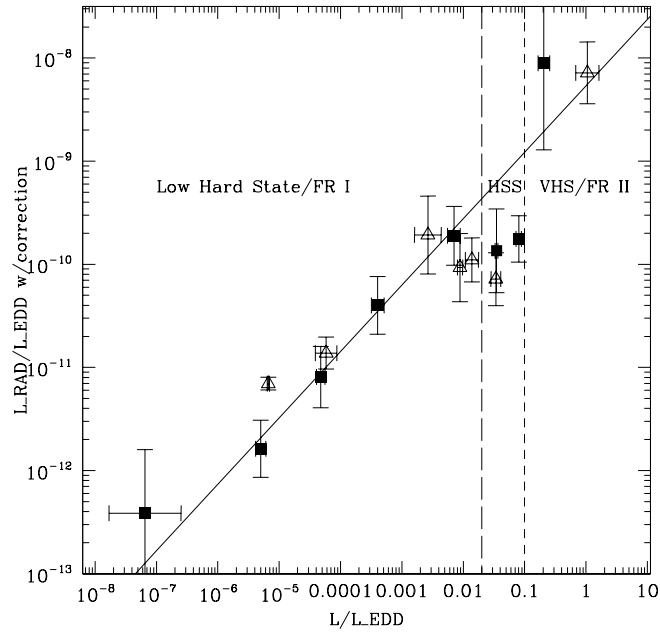,width=9cm,angle=0}}
\caption{
Jet power as a function of Eddington-scaled luminosity for X-ray
binaries ({\it open triangles}) and AGN ({\it solid squares}), adjusted for the
mass term of Merloni, Heinz \& di Matteo (2003); from Maccarone, Gallo
\& Fender (2003). The reduction in radio power in the
approximate range 1\%--10\% Eddington is well-known in X-ray binaries as
the quenching of the jet in soft X-ray states, a phenomenon that may also therefore occur in AGN. The data for GRS 1915+105 lie
in the most luminous bin, corresponding to FRII-type AGN.}
\label{}
\end{figure}

%Figure 26
\begin{figure}
\centerline{\psfig{figure=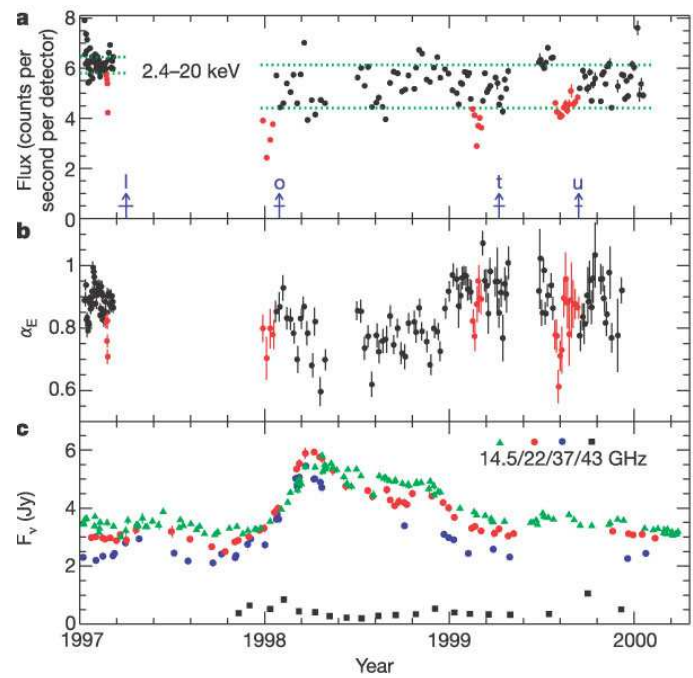,width=9cm,angle=0}}
\caption{Possible analogous disc-jet coupling in the AGN 3C 120, from
  Marscher et al. 2002}
\label{}

\end{figure}

%Figure 27
\begin{figure}
\centerline{\psfig{figure=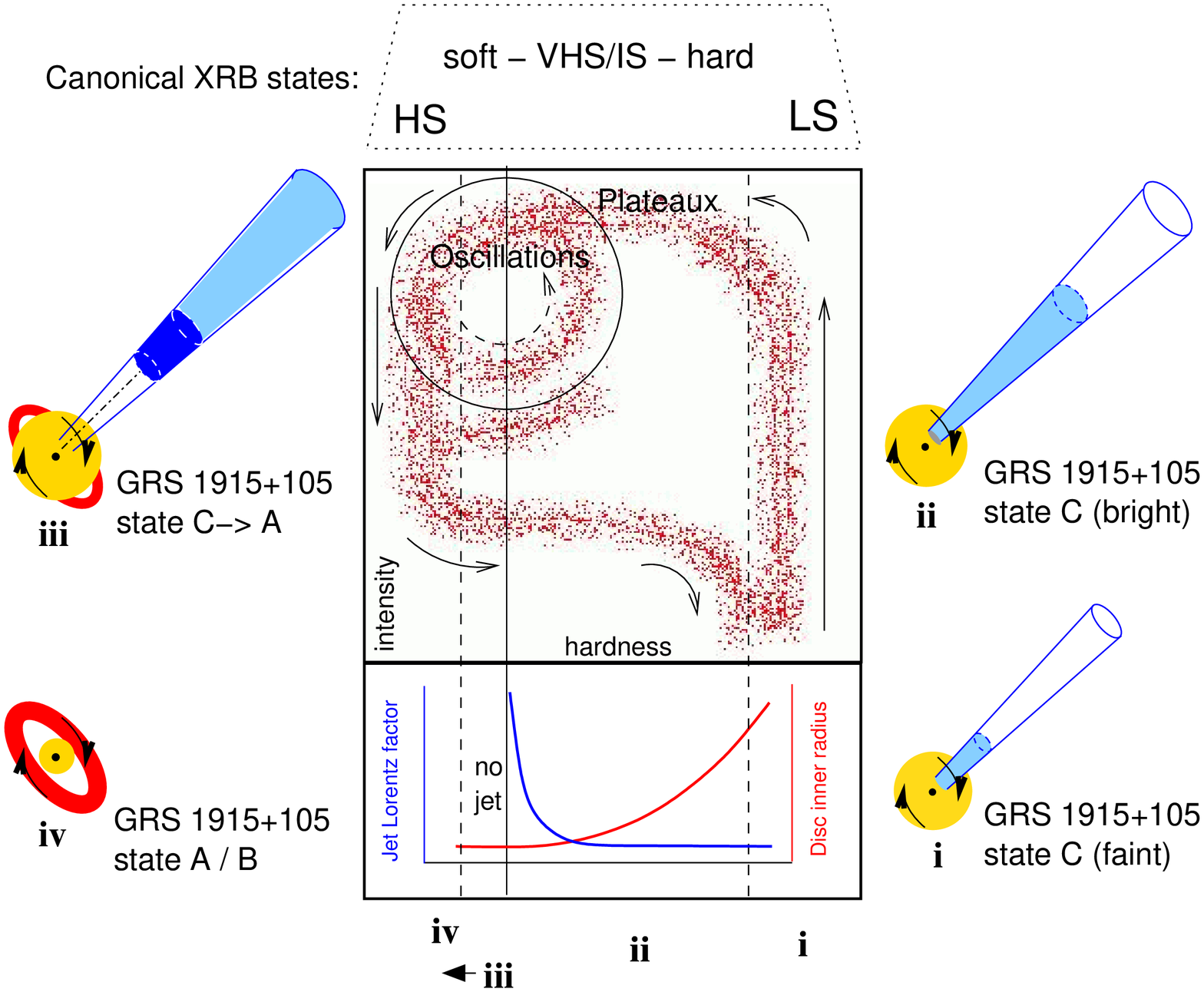, angle=0,width=15cm}}
\caption{A schematic illustrating our model for the evolution of
outbursts and related phases of jet formation in GRS 1915+105 and
other accreting black holes. Typical X-ray binary outbursts follow the
path of the solid arrows, transiting from the canonical LS to the
canonical HS via a monotonically softening VHS/IS. This evolution
corresponds to leftward motion across the hardness-intensity diagram
({\it top panel}). We suggest that above some hardness a jet is
produced, below that hardness the jet is suppressed, and that close to
the transition (indicated by the vertical line) the jet velocity
increases to significantly relativistic values ({\it lower panel}).
This line may correspond to the geometrically thin accretion disc
reaching the innermost stable circular orbit around the black hole.As
a result, crossing the vertical line from left to right results in a
shock in the outflow and subsequent cessation of core jet production;
this is observed as a singular optically thin radio event. GRS
1915+105 seems to be semicontinuously located in the upper-left region
of the diagram, repeatedly crossing the vertical line, producing
optically thin radio flares each time it does so. A loop, such as that
indicated by the circle, may correspond to a single oscillation
event. Extended plateau phases correspond to excursions towards the
canonical LS, but are inevitably terminated by a softening and major
radio event as the source crosses the vertical line once more. This
model seems to be consistent with the behavior of other black hole
XRBs, and may even be applicable to AGN.}
\label{both}
\end{figure}


\begin{thebibliography}{}

\bibitem[]{abra03}
Abramowicz MA, Karas V, Kluzniak W, Lee WH, Rebusco P. 2003.
	{\it Publ. ASJ} 55:467--71

\bibitem[]{alexan94} 
Alexandrovich N, Borozdin K, Sunyaev R. 1994. {\it IAU Circ. No. 6080}

\bibitem[]{}
Atoyan AM, Aharonian FA. 1999. {\it MNRAS} 302:253--76

\bibitem[]{}
Belczynski K, Bulik T. 2002. {\it Ap. J.} 574:L147--50

\bibitem[]{belloni1999a}
Belloni T. 1999a. {\it Astrophys. Lett. Comm.} 38:225--28

\bibitem[]{belloni1999b}
Belloni T. 1999b. {\it MPE Rep.} 272:82--85

\bibitem[]{belloni2001}
Belloni T. 2001. In {\it The Neutron Star---Black Hole Connection, 
	NATO ASI, Elounda,} C567:295--300. Dordrecht: Kluwer 

\bibitem[]{belloni00} 
Belloni T, Klein-Wolt M, M\'endez M, van der Klis M, van Paradijs J.
	2000. {\it Astron. Astrophys.} 355:271--90

\bibitem[]{belloni97a} 
Belloni T, M\'endez M, King AR, van der Klis M, van Paradijs J.
	1997a. {\it Ap. J. Lett.} 479:145--48

\bibitem[]{belloni97b} 
Belloni T, M\'endez M, King AR, van der Klis M, van Paradijs J.
	1997b. {\it Ap. J. Lett.} 488:109--12

\bibitem[]{belal2001}
Belloni T, M\'endez M, S\'anchez-Fern\'andez C. 2001. 
	{\it Astron. Astrophys.} 372:551--56


\bibitem[]{belloni02} 
Belloni T, Nespoli E, Homan J, van der Klis M, Lewin WHG, et al. 2002.
	{\it New Views on Microquasars}, pp. 83--85. Kolkata, India: Cent. Space
	Phys. 

\bibitem[]{}
Blandford RD, K\"onigl A. 1979. {\it Ap. J.} 232:34--48

\bibitem[]{}
Bodo G, Ghisellini G. 1995. {\it Ap. J.} 441:L69--71

\bibitem[]{}
Bo\"er M, Greiner J, Motch C. 1996. {\it Astron. Astrophys.} 305:835--38

\bibitem[]{}
Brocksopp C, Fender RP, McCollough M, Pooley GG, Rupen MP, et al. 2002. {\it MNRAS} 331:765--75

\bibitem[]{}
Burbidge GR. 1959. {\it Ap. J.} 12

\bibitem[]{castro92} 
Castro-Tirado AJ, Brandt S, Lund N. 1992. {\it IAU Circ. No. 5590}

\bibitem[]{castro94} 
Castro-Tirado AJ, Brandt S, Lund N, Laphsov I, Sunyaev RA, et al. 1994.
	{\it Ap. J. Suppl.} 92:469--72

\bibitem[]{castro93} 
Castro-Tirado AJ, Davies J, Brandt S, Lund N. 1993.
	{\it IAU Circ. No. 5830}

\bibitem[]{}
Castro-Tirado AJ, Geballe TR, Lund N. 1996. {\it ApJ} 461:L99--101


\bibitem[]{}
Chaty S, Mirabel IF, Duc PA, Wink JE, Rodr\'iguez LF. 1996. {\it
  Astron. Astrophys.} 310:825--30

\bibitem[]{}
Chaty S, Rodr\'iguez LF, Mirabel IF, Geballe TR, Fuchs Y, et al. 2001. {\it Astron. Astrophys}. 366:1035--46

\bibitem[]{chen97} 
Chen X, Swank JH, Taam RE. 1997. {\it Ap. J. Lett.} 477:41--44

\bibitem[]{}
Collins RS, Kaise CR, Cox SJ. 2003. {\it MNRAS} 338:1365--87

\bibitem[]{}
Corbel S, Kaaret P, Jain RK, Bailyn CD, Fender RP, et al. 2001. {\it Ap. J}. 554:43--48

\bibitem[]{}
Corbel S, Nowak MA, Fender RP, Tzioumis AK, Markoff S. 2003. {\it Astron. Astrophys} 400:1007--12

\bibitem[]{cui99}
Cui W, Zhang SN, Chen W, Morgan EH. 1999. {\it Ap. J. Lett.} 512:43--46

\bibitem[]{}
Dhawan V, Mirabel IF, Rodr\'iguez LF. {\it Ap. J.} 543:373--85


\bibitem[]{}
Eikenberry SS, Fazio GG. 1997. {\it Ap. J.} 476:281--90

\bibitem[]{}
Eikenberry SS, Matthews K, Morgan EH, Remillard RA, Nelson
RW. 1998a. {\it Ap. J.} 494:L61--64

\bibitem[]{}
Eikenberry SS, Matthews K, Muno M, Blanco PR, Morgan EH, Remillard
RA. 2000. {\it Ap. J.} 532:L33--36

\bibitem[]{}
Eikenberry SS, Matthews K, Murphy TW Jr, Nelson RW, Morgan EH, et al.
1998b. {\it Ap. J.} 506:L31--34


\bibitem[]{esi97}
Esin AA, McClintock JE, Narayan R. 1997. {\it Ap. J.} 489:865--89

\bibitem[]{}
Falcke H, Biermann PL. 1996. {\it Astron. Astrophys.} 308:321--29

\bibitem[]{}
Falcke H, Koerding E, Markoff S. 2003. {\it Astron. Astrophys.} In press (astro-ph/0305335)

\bibitem[]{}
Fender RP. 2001. {\it MNRAS} 322:31--42

\bibitem[]{}
Fender RP. 2003. {\it MNRAS} 340:1353--58

\bibitem[]{fender94} 
Fender RP. 2004. In {\it Compact Stellar X-ray Sources.} Cambridge, UK:
	Cambridge Univ. Press. In press (astro-ph/0303339)

\bibitem[]{}
Fender RP, Belloni TM, Gallo E. 2004b. {\it MNRAS} Submitted

\bibitem[]{}
Fender RP, Gallo E, Jonker PG. 2003. {\it MNRAS} 343:L99--103

\bibitem[]{} 
Fender RP, Garrington ST, McKay DJ, Muxlow TWB, Pooley
GG, et al. 1999. {\it MNRAS} 304:865--76

\bibitem[]{}
Fender RP, Hjellming RM, Tilanus RPJ, Pooley GG,Deane JR,
et al. 2001. {\it MNRAS} 322:L23--27

\bibitem[]{}
Fender RP, Kuulkers E. 2001. {\it MNRAS} 324:923--30

\bibitem[]{}
Fender RP, Pooley GG. 1998. {\it MNRAS} 300:573--76

\bibitem[]{}
Fender RP, Pooley GG. 2000. {\it MNRAS} 318:L1--5

\bibitem[]{}
Fender RP, Pooley GG, Brocksopp C, Newell SJ. 1997. {\it MNRAS} 
290:L65--69

\bibitem[]{}
Fender RP, Rayner D, Trushkin SA, O'Brien K, Sault RJ, et al.
2002a. {\it MNRAS} 330:212--18

\bibitem[]{}
Fender RP, Rayner D, McCormick DG, Muxlow TMB, Pooley GG, et al.
2002b. {\it MNRAS} 336:39--46

\bibitem[]{}
Fender RP, Wu K, Johnston H, Tzioumis T, Jonker P, et al. 2004a. {\it Nature} 427:222--24


\bibitem[]{feroci99}
Feroci M, Matt G, Pooley G, Costa E, Tavani M, Belloni T. 1999.
	{\it Astron. Astrophys.} 351:985--92

\bibitem[]{}
Foster RS, Waltman EB, Tavani M, Harmon BA, Zhang SN, et al. 1996.
{\it ApJ} 467:L81--84

\bibitem[]{}
Fuchs Y, Mirabel IF, Claret A. 2003a. {\it Astron. Astrophys.} 404:1011--21

\bibitem[]{fuchs03}
Fuchs Y, Rodriguez J, Mirabel IF, Chaty S, Rib\'o M, et al. 2003b.
	{\it Astron. Astrophys.} 409:L35--39

\bibitem[]{}
Gallo E, Corbel S, Fender RP, Maccarone TJ, Tzioumis
AK. 2004. {\it MNRAS} 347:L52

\bibitem[]{}
Gallo E, Fender R, Pooley G. 2003. {\it MNRAS} 344:60--72

\bibitem[]{}
Gerard E, Rodr\'iguez LF, Mirabel IF. 1994. {\it IAU Circ. No. 5958}

\bibitem[]{}
Giovannini G, Feretti L, Tordi M, Venturi T, Massaglia S, et al. 2001. {\it Astrophys. Space Sci. Suppl.} 276:111--12

\bibitem[]{green03} 
Greenhough J, Chapman SC, Chaty S, Dendy RO, Rowlands G. 2003. 
	{\it MNRAS} 340:851--55

\bibitem[]{greiner93} 
Greiner J. 1993. {\it IAU Circ. No. 5786}

\bibitem[]{}
Greiner J, Cuby JG, McCaughrean MJ, Castro-Tirado AJ, Mennickent
RE. 2001a. {\it Astron. Astrophys.} 373:L37--40

\bibitem[]{}
Greiner J, Cuby JG, McCaughrean MJ. 2001b. {\it Nature} 414:522--524

\bibitem[]{greiner96} 
Greiner J, Morgan EH, Remillard RA. 1996. {\it Ap. J. Lett.} 473:107--10


\bibitem[]{grove98}
Grove JE, Johnson WN, Kroeger RA, McNaron-Brown K, Skibo JG, Phlips BF.
	1998. {\it Ap. J.} 500:899--908

\bibitem[]{diana03}
Hannikainen DC, Vilhu O, Rodriguez J, Brandt S, Westergaard NJ, et al. 2003.
	{\it Astron. Astrophys.} 411:L451

\bibitem[]{harmon92} 
Harmon BA, Paciesas WS, Fishman GJ. 1992. {\it IAU Circ. No. 5619}

\bibitem[]{harmon95} 
Harmon BA, Deal KJ, Paciesas WS, Zhang SN, Robinson CR, et al.
	1997. {\it Ap. J. Lett.} 477:85--89

\bibitem[]{}
Heinz S. 2002. {\it Astron. Astrophys.} 388:L40--43

\bibitem[]{}
Hjellming RM, Han H. 1995.  In
        {\it X-ray Binaries,} p. 308. Cambridge, UK: Cambridge Univ. Press 

\bibitem[]{}
Hjellming RM, Johnston KJ. 1988. {\it Ap. J.}. 328:600--9

 
\bibitem[]{hjel95} 
Hjellming RM, Rupen MP. 1995. {\it Nature} 375:464--68


\bibitem[]{}
Homan J, Wijnands R, van der Klis M, Belloni T, van Paradijs J, et al. 2001. {\it Ap. J. Supp. Ser.} 132:377--402

\bibitem[]{jaho96} 
Jahoda K, Swank JH, Giles AB, Stark MJ, Strohmayer T, et al.
1996. {\it Proc. SPIE} 2808:59--70

\bibitem[]{janiuk98}
Janiuk A, Czerny B, Siemiginoswka A. 2000a. {\it Ap. J. Lett.} 542:33--36

\bibitem[]{janiuk01}
Janiuk A, Czerny B, Siemiginoswka A. 2000b. {\it Ap. J.} 576:908--22

\bibitem[]{ji03}
Ji JF, Zhang SN, Qu JL, Li TP. 2003. {\it Ap. J. Lett.} 584:23--26

\bibitem[]{}
Jorstad SG, Marscher AP, Mattox JR, Wehrle AE, Bloom SD,
Yurchenko AV. 2001. {\it Ap. J. Suppl. Ser.} 134:181--240



\bibitem[]{}
Kaiser CR, Gunn KE, Brocksopp C, Sokoloski JL. 2004. {\it MNRAS}. Submitted

\bibitem[]{}
Kaiser CR, Sunyaev R, Spruit HC. 2000 {\it Astron. Astrophys.} 356:975--88

\bibitem[]{}
King AR. 2004. {\it MNRAS} 347:L18--20

\bibitem[]{klein03}
Klein-Wolt M, Fender RP, Pooley GG, Belloni T, Migliari S. 2002.
	{\it MNRAS} 331:745--64

\bibitem[]{kotani00}
Kotani T, Ebisawa K, Dotani T, Inoue H, Nagase F, et al. 2000.
	{\it Ap. J.} 539:413--23

\bibitem[]{}
Krolik J.H., `Active Galactic Nuclei: from the central black hole to
the galactic environment', Princeton Series in Astrophysics,
Princeton, New Jersey, 1999

\bibitem[]{lee02}
Lee JC, Reynolds CS, Remillard R, Schulz NS, Blackman EG, Fabian AC.
	2002. {\it Ap. J.} 567:1102--11

\bibitem[]{levine96} 
Levine AM, Bradt H, Cui W, Jernigan JG, Morgan EH, et al.
	1996. {\it Ap. J. Lett.} 469:33--36

\bibitem[]{}
Levinson A, Blandford RD. 1996. {\it Ap. J.} 456:L29--33

\bibitem[]{lin00}
Lin D, Smith IA, Liang EP, B\"ottcher M. 2000. {\it Ap. J. Lett.} 543:141--44

\bibitem[]{livio03}
Livio M, Pringle JE, King AR. 2003. {\it Ap. J.} 593:184--88

\bibitem[]{}
Maccarone TJ. 2002. {\it MNRAS} 336:1371--76

\bibitem[]{}
Maccarone TJ, Coppi PS. 2003. {\it MNRAS} 338:189--96

\bibitem[]{}
Maccarone TJ, Gallo E, Fender R. 2003. {\it MNRAS} 345:L19--24

\bibitem[]{}
Malzac J, Merloni A, Fabian AC. 2004. {\it MNRAS}. In press

\bibitem[]{}
Markoff S, Falcke H, Fender RP. 2001. {\it Astron. Astrophys.} 372:L25--28

\bibitem[]{markw99} 
Markwardt CB, Swank JH, Taam RE. 1999. {\it Ap. J. Lett.} 513:37--40

\bibitem[]{}
Marscher AP, Jorstad SG, Gomez JL, Aller MF, Terasranta H, et al. 2002. {\it Nature} 417:625--27

\bibitem[]{marto02}
Martocchia A, Matt G, Karas V, Belloni T, Feroci M. 2002.
	{\it Astron. Astrophys.} 387:215--21

\bibitem[]{mcc03}
McClintock JE, Remillard RA. 2004.
In {\it Compact Stellar X-ray Sources.} Cambridge, UK:
        Cambridge Univ. Press. In press (astro-ph/0306213)

\bibitem[]{}
Meier D. 1999. {\it Ap. J.} 522:753--66

\bibitem[]{}
Meier D. 2001. {\it Ap. J.} 548:L9--12

\bibitem[]{}
Meier DL, Koide S, Uchida Y. 2001. {\it Science} 291:84--92

\bibitem[]{merl2000} 
Merloni A, Fabian AC, Ross RR. 2000. {\it MNRAS} 313:193--97

\bibitem[]{}
Merloni A, Heinz S, di Matteo T. 2003. {\it MNRAS} 345:1057--76

\bibitem[]{migliari03} 
Migliari S, Belloni T. 2003. {\it Astron. Astrophys.} 404:283--89

\bibitem[]{}
Mirabel IF, Bandyopadhyay R, Charles PA, Shahbaz T, Rodr\'iguez
LF. 1997. {\it Ap. J.} 477:L45--48

\bibitem[]{}
Mirabel IF, Dhawan V, Chaty S, Rodr\'iguez LF, Marti J, et al.
1998. {\it Astron. Astrophys} 333:L1--4

\bibitem[]{mira93b} 
Mirabel IF, Duc PA, Teyssier R, Paul J, Rodr\'iguez LF, et al.
	1993b. {\it IAU Circ. No. 5830}

\bibitem[]{mira94} 
Mirabel IF, Duc PA, Rodr\'iguez LF, Teyssier R, Paul J, et al.
	1994. {\it Astron. Astrophys. Lett.} 282:17--20

\bibitem[]{mira94} 
Mirabel IF, Rodr\'iguez LF. 1994. {\it Nature} 371:46--48

\bibitem[]{}
Mirabel IF, Rodr\'iguez LF. 1999. {\it Annu. Rev. Astron. Astrophys.} 37:409--43

\bibitem[]{}
Mirabel IF, Rodr\'iguez LF, Cordier B, Paul J, Lebrun F. 1992. {\it
  Nature} 358:215--17


\bibitem[]{mira93a} 
Mirabel IF, Rodr\'iguez LF, Marti J, Teyssier R, Paul J, Auriere M.
	1993a. {\it IAU Circ. No. 5773}

\bibitem[]{}
Mirabel IF, Rodr\'iguez LF, Chary S, Sauvage M, Gerard E, et al. 1996. {\it Ap. J.} 472:L111--14


\bibitem[]{mitsuda84} 
Mitsuda K, Inoue H, Koyama K, Makishima K, Matsuoka M, et al.
	1984. {\it Publ. ASJ} 36:741--59

\bibitem[]{mgr97}
Morgan EH, Remillard RA, Greiner J. 1997. {\it Ap. J.} 482:993--1010

\bibitem[]{muno99}
Muno MP, Morgan EH, Remillard RA. 1999. {\it Ap. J.} 527:321--40

\bibitem[]{muno01}
Muno MP, Remillard RA, Morgan EH, Waltman EB, Dhawan V, et al. 2001. 
	{\it Ap. J.} 556:515--32

\bibitem[]{}
Naik S, Agrawal PC, Rao AR, Paul B, Seetha S, Kasturirangan
K. 2001. {\it Ap. J}. 546:1075--85

\bibitem[]{naik02}
Naik S, Agrawal PC, Rao AR, Paul B. 2002. {\it MNRAS} 330:487--96

\bibitem[]{}
Naik S, Rao AR. 2000. {\it Astron. Astrophys.}. 362:691--96

\bibitem[]{nandi2001}
Nandi A, Chakrabarti SK, Vadawale SV, Rao AR. 2001. {\it Astron. Astrophys.}
	380:245--50

\bibitem[]{naya00}
Nayakshin S, Rappaport S, Melia F. 2000. {\it Ap. J.} 535:798--814

\bibitem[]{nes03}
Nespoli E, Belloni T, Homan J, Miller JM, Lewin WHG, et al. 2003.
{\it Astron. Astrophys.}  412:235

\bibitem[]{}
Ogley RN, Bell B, Fender RP, Pooley GG, Waltman EB. 2000. {\it MNRAS} 317:158--62

\bibitem[]{}
Ostrowski M, F\"urst E. 2001. {\it Astron. Astrophys.} 367:613--16

\bibitem[]{pac96} 
Paciesas WS, Deal KJ, Harmon BA, Zhang SN, Wilson CA, et al.
	1996. {\it Astron. Astrophys. Suppl.} 120:205--208

\bibitem[]{paul97} 
Paul B, Agrawal PC, Rao AR, Vahia MN, Yadav JS, et al.
	1997. {\it Astron. Astrophys. Lett.} 320:37--40

\bibitem[]{}
Podsiadlowski P, Rappaport S, Han Z. 2003. {\it MNRAS} 341:385--404

\bibitem[]{}
Pooley GG. 1995. {\it IAU Circ. No. 6269}

\bibitem[]{}
Pooley GG, Fender RP. 1997. {\it MNRAS} 292:925--33

\bibitem[]{rau2003}
Rau A, Greiner J. 2003. {\it Astron. Astrophys.} 397:711--22

\bibitem[]{raual2003}
Rau A, Greiner J, McCollough ML. 2003. {\it Ap. J. Lett.} 590:37--40

\bibitem[]{}
Rees MJ. 1978. {\it MNRAS} 184:P61--65

\bibitem[]{}
Rees MJ, Meszaros P. 1994. {\it Ap. J.} 430:L93--96

\bibitem[]{reig2003}
Reig P, Belloni T, van der Klis M. 2003. {\it Astron. Astrophys.} 412:229

\bibitem[]{reig2000}
Reig P, Belloni T, van der Klis M, M\'endez M, Kylafis ND, Ford EC. 2000.
	{\it Ap. J.} 541:883--88

\bibitem[]{rodrig2002a}
Rodriguez J, Durouchoux P, Mirabel IF, Ueda Y, Tagger M, Yamaoka K. 2002a.
	{\it Astron. Astrophys.} 386:271--79

\bibitem[]{rodrig2002b}
Rodriguez J, Varni\'ere P, Tagger M, Durouchoux P. 2002b.
	{\it Astron. Astrophys.} 387:487--96

\bibitem[]{rodrig95} 
Rodr\'iguez LF, Gerard E, Mirabel IF, Gomez Y, Velasquez A. 
	1995. {\it Ap. J. Suppl.} 101:173--79

\bibitem[]{rodrig93} 
Rodr\'iguez LF, Mirabel IF.  1993. {\it IAU Circ. No. 5900}

\bibitem[]{}
Rodr\'iguez LF, Mirabel IF. 1997. {\it Ap. J} 474:L123--25

\bibitem[]{}
Rodr\'iguez LF, Mirabel IF. 1999. {\it Ap. J} 511:398--404


\bibitem[]{roth98} 
Rothschild RE, Blanco PR, Gruber DE, Heindl WA, MacDonald DR, et al.
	1998. {\it Ap. J.} 496:538--49

\bibitem[]{}
Sams BJ, Eckart A, Sunyaev R. 1996a. {\it Nature} 382:47 

\bibitem[]{}
Sams BJ, Eckart A, Sunyaev R. 1996b. {\it IAU Circ. No. 6455}

\bibitem[]{sazo95} 
Sazonov S, Sunyaev R.  1995. {\it IAU Circ. No. 6209}

\bibitem[]{}
Spada M, Ghisellini G, Lazzati D, Celotti A. 2001. {\it MNRAS} 325:1559--70

\bibitem[]{}
Stirling AM, Spencer RE, de la Force CJ, Garrett MA, Fender RP, Ogley RN. 2001. {\it MNRAS} 327:1273--78

\bibitem[]{stro01}
Strohmayer TE. 2001. {\it Ap. J. Lett.} 554:169--72

\bibitem[]{}
Sunyaev RA, Titarchuk LG. 1980. {\it Astron. Astrophys.} 86:121--38

\bibitem[]{ewa98}
Szuszkiewicz E, Miller JC. 1998. {\it MNRAS} 298:888--96

\bibitem[]{ewa01}
Szuszkiewicz E, Miller JC. 2001. {\it MNRAS} 328:36--44

\bibitem[]{taam97}
Taam RE, Chen X, Swank JH. 1997. {\it Ap. J. Lett.} 485:83--86

\bibitem[]{tanlew95} 
Tanaka Y, Lewin WHG. 1995. In 
	{\it X-ray Binaries,} p. 126. Cambridge, UK: Cambridge Univ. Press. 

\bibitem[]{}
Tingay SJ, Jauncey DL, Preston RA, Reynolds JE, Meier DL, et al. 1995. {\it Nature} 374:141--43

\bibitem[]{tomkaa01}
Tomsick JA, Kaaret P. 2001. {\it Ap. J.} 548:401--9

\bibitem[]{trudo2001}
Trudolyubov SP. 2001. {\it Ap. J.} 558:276--82

\bibitem[]{trudo1999a}
Trudolyubov SP, Churazov EM, Gilfanov MR. 1999a. {\it Astron. Lett.}
	25:718--38

\bibitem[]{trudo1999b}
Trudolyubov SP, Churazov EM, Gilfanov MR. 1999b. {\it Astron. Astrophys. Lett.}
	351:15--18

\bibitem[]{vdk95}
van der Klis M. 1995. In
        {\it X-ray Binaries,} p. 252. Cambridge, UK: Cambridge Univ. Press 

\bibitem[]{}
van der Laan H. 1996. {\it Nature} 211:1131--33

\bibitem[]{}
Vadawale SV, Rao AR, Chakrabarti SK. 2001. {\it Astron. Astrophys.} 372:793--802

\bibitem[]{}
Vadawale SV, Rao AR, Naik S, Yadav JS, Ishwara-Chandra CH,
et al. 2003. {\it Ap. J.} 597:1023--35

\bibitem[]{vignarca}
Vignarca F, Migliari S, Belloni T, Psaltis D, van der Klis M. 2003.
	{\it Astron. Astrophys.} 397:729--38

\bibitem[]{vilhu98} 
Vilhu O, Nevalainen J. 1998. {\it Ap. J. Lett.} 508:85--89

\bibitem[]{vilhu01} 
Vilhu O, Poutanen J, Nikula P, Nevalainen J. 2001. 
	{\it Ap. J. Lett.} 553:51--54
	
\bibitem[]{}
Winkler C, Trams N. 1998. {\it Astron. Astrophys.} 337:729--38

\bibitem[]{yadav99} 
Yadav JS, Rao AR, Agrawal PC, Paul B, Seetha S, et al. 1999.
	{\it Ap. J.} 517:935--50

\bibitem[]{}
Yu W, Klein-Wolt M, Fender R, van der Klis M. 2003. {\it Ap. J.} 589:L33-36

\bibitem[]{zamp2001}
Zampieri L, Turolla R, Szuszkiewicz E. 2001. {\it MNRAS} 325:1266--74

\bibitem[]{zdz2001}
Zdziarski AA, Grove JE, Poutanen J, Rao AR, Vadawale SV. 2001.
	{\it Ap. J. Lett.} 554:45--48

\bibitem[]{}
Zdziarski AA, Lubinski P, Gilfanov G, Revnivtsev M. 2003. {\it
  MNRAS} 342:355--72

\bibitem[]{zhang95} 
Zhang SN, Harmon BA, Paciesas WS, Fishman GJ.
	1995. {\it IAU Circ. No. 6209}


\end{thebibliography}
\end{document}